\documentclass[notitlepage,10pt,aps,prd,twocolumn,amsmath,amssymb,floatfix,superscriptaddress,nofootinbib,preprintnumbers]{revtex4-1}

\usepackage{amsfonts,amssymb,amsmath,graphicx,color,bm,epsfig}
\usepackage[normalem]{ulem}
\usepackage{multirow}
\usepackage{centernot}

\definecolor{ultramarine}{rgb}{0.07, 0.04, 0.56}
\definecolor{cadmiumgreen}{rgb}{0.0, 0.42, 0.24}
\definecolor{indigo(dye)}{rgb}{0.0, 0.25, 0.42}
\usepackage[linktocpage=true]{hyperref}
\hypersetup{
colorlinks=true,
citecolor=ultramarine,
linkcolor=cadmiumgreen,
urlcolor=indigo(dye),
}

\newcommand{\Mpl}{M_{\rm Pl}}

\newcommand{\e}{\epsilon_H}
\renewcommand{\O}{\mathcal{O}}

\newcommand{\N}{\mathcal{N}}
\newcommand{\pa}{\partial}

\newcommand{\GeV}{{\rm \ GeV}}
\newcommand{\hcr}{h_{\rm max}}
\newcommand{\hcl}{h_{\rm cl}}
\newcommand{\hend}{h_{\rm end}}
\newcommand{\hrescue}{h_{\rm rescue}}

\newcommand{\SM}{{\rm SM}}
\newcommand{\BSM}{{\rm BSM}}

\newcommand{\VT}{V^T}
\newcommand{\widebar}[1]{\mkern 1.5mu\overline{\mkern-1.5mu#1\mkern-1.5mu}\mkern 1.5mu}
\newcommand{\medbar}[1]{\mkern 3mu\overline{\mkern-3mu#1\mkern-3mu}\mkern 3mu}

\newcommand\numberthis{\addtocounter{equation}{1}\tag{\theequation}}
\definecolor{darkgreen}{cmyk}{0.85,0.2,1.00,0.2} 
\definecolor{purple}{cmyk}{0.5,1.0,0,0} 

\begin{document}

\preprint{YITP-19-51}

\title[]{Primordial black holes as dark matter through Higgs field criticality}

\author{Samuel Passaglia}
\email{passaglia@uchicago.edu}
\affiliation{Kavli Institute for Cosmological Physics, Department of Astronomy \& Astrophysics, 
Enrico Fermi Institute, University of Chicago, Chicago, IL 60637}

\author{Wayne Hu}
\affiliation{Kavli Institute for Cosmological Physics, Department of Astronomy \& Astrophysics, 
Enrico Fermi Institute, University of Chicago, Chicago, IL 60637}

\author{Hayato Motohashi}
\thanks{Present address: Division of Liberal Arts, Kogakuin University, 2665-1 Nakano-machi, Hachioji, Tokyo, 192-0015, Japan}
\affiliation{Center for Gravitational Physics, Yukawa Institute for Theoretical Physics, Kyoto University, Kyoto 606-8502, Japan}

\label{firstpage}

\begin{abstract}
We study the dynamics of a spectator Higgs field which stochastically evolves during inflation onto near-critical trajectories on the edge of a runaway instability. We show that its fluctuations do not produce primordial black holes (PBHs) in sufficient abundance to be the dark matter, nor do they produce significant second-order gravitational waves. First we show that the Higgs produces larger fluctuations on cosmic microwave background scales than on PBH scales, itself a no-go for a viable PBH scenario. Then we track the superhorizon perturbations nonlinearly through reheating using the $\delta N$ formalism to show that they are not converted to large curvature fluctuations.  Our conclusions hold regardless of any fine-tuning of the Higgs field for both the Standard Model Higgs and for Higgs potentials modified to prevent unbounded runaway.
\end{abstract}

\date{\today}

\maketitle 

\section{Introduction}

Bottom-up naturalness suggests that new physics should appear at the TeV scale to stabilize the Higgs mass \cite{Veltman:1980mj}. Unfortunately, no physics beyond the Standard Model (SM) has yet been found by the LHC, nor have weakly interacting massive particles been found in dark matter direct detection experiments. 

If naturalness is abandoned as a guiding principle, then there may be no new physics intervening before the Planck scale. If this possibility is taken seriously, then the SM exhibits the unusual property that the Higgs effective potential turns over at large field values \cite{Degrassi:2012ry} and our electroweak vacuum is metastable, with a lifetime longer than the age of our Universe \cite{Andreassen:2017rzq}.

This near criticality of the Higgs has cosmological consequences \cite{Espinosa:2007qp,EliasMiro:2011aa,Espinosa:2015qea,Markkanen:2018pdo}, and it may also be that cosmology can help elucidate its origins and provide the solution to the Higgs naturalness problem \cite{Buttazzo:2013uya,Khoury:2019yoo,Giudice:2019iwl}.

In particular, Ref.~\cite{Espinosa:2017sgp} proposed that Higgs criticality could explain the existence of the dark matter within the SM as primordial black holes (PBHs).  Furthermore this scenario provides a potential anthropic reason why the Higgs should be near critical \cite{Gross:2018ivp,Espinosa:2018euj}.
Here the Higgs field is a spectator during inflation and not the inflaton itself if its coupling to gravity is not large (see, e.g., \cite{Hamada:2014wna,Ezquiaga:2017fvi,Salvio:2018rv,Masina:2018ejw}). 
Large Higgs fluctuations are formed during inflation, and they are potentially converted to large superhorizon curvature perturbations after inflation due to Higgs criticality.
Once they reenter the horizon during radiation domination, PBHs of horizon scale mass form from
their gravitational collapse (see Ref.~\cite{Sasaki:2018dmp} for a recent review) and in certain specific mass
ranges can comprise the entire dark matter abundance \cite{Niikura:2017zjd,Montero-Camacho:2019jte}.

To form PBHs in sufficient abundance to be the dark matter, curvature perturbations must be
amplified to $\sim 10^{-1}$ on small scales while not violating the $\sim 10^{-5}$ constraints on cosmic microwave background (CMB) scales \cite{Motohashi:2017kbs,Passaglia:2018ixg}.
We will focus in this paper on both these conditions and carefully assess the viability of this scenario.

This paper is organized as follows: in \S\ref{sec:mechanism} we provide a general overview of the Higgs instability and the PBH production mechanism; in \S\ref{sec:inflation} we compute the Higgs power spectrum produced during inflation on all observationally relevant scales; in \S\ref{sec:reheating} we track these Higgs fluctuations nonlinearly through reheating to compute the curvature fluctuations on which the PBH abundance depends; and we summarize our results in \S\ref{sec:conclusions}.

Throughout this paper, a prime\, $'$\, denotes a derivative with respect to the $e$-folds $N$, where $N=0$ marks the end of inflation, and an overdot\, $\dot{}$\,  
denotes a derivative with respect to the conformal time $\eta$. We assume a spatially flat background metric throughout and work in units in which $\Mpl\equiv 1/\sqrt{8\pi G}=1$ unless otherwise specified.

\section{Higgs Instability Mechanism}
\label{sec:mechanism}

In this section, we review the general features of the Higgs instability mechanism for producing PBH dark matter and the principles governing the spectrum of perturbations it generates.

The Higgs field acts as a spectator during inflation, which is driven by an inflaton field, but converts its quantum fluctuations into curvature fluctuations once it decays after inflation. If these quantum fluctuations are amplified into large enough curvature fluctuations  by the Higgs instability, they will form PBHs when they reenter the horizon in the radiation dominated epoch.

In particular, after inflation the spatial metric on a uniform total energy density hypersurface,
\begin{equation}
\label{eq:sign_convention}
g_{ij} = a^2 e^{2 \zeta} \delta_{i j},
\end{equation}
 possesses a  curvature perturbation $\zeta$ that can be decomposed as
\begin{equation} 
\label{eq:zeta_tot_intro}
\zeta =\left(1-   \frac{\rho_h'}{\rho_{\rm tot}'} \right)  \zeta_{\rm r} + \frac{\rho_h'}{\rho_{\rm tot}'} \zeta_h,
\end{equation}
in linear theory.  Here
$\zeta_{\rm r}$ is the curvature perturbation on hypersurfaces of uniform energy density $\rho_{\rm r}$ of reheat products from the inflaton, and $\zeta_h$ is the curvature perturbation on hypersurfaces of uniform Higgs energy density $\rho_h$.
Here $\rho_{\rm tot} \equiv \rho_{\rm r} + \rho_h$ is the total energy density.

For this mechanism to succeed observationally, the variance per logarithmic interval in $k$ of the Fourier mode function $\zeta^k$, 
\begin{equation}
\Delta^2_{\zeta}(k)\equiv \frac{k^3}{2\pi^2}|\zeta^k|^2,
\end{equation}
must be small on the large scales probed by anisotropies in the CMB, 
\begin{equation}
\label{eq:delta_2_cmb}
\Delta^2_{\zeta}(k_{\rm CMB}) \sim 10^{-9},
\end{equation}
while on small scales associated with primordial black holes the power spectrum must reach
\begin{equation}
\label{eq:delta_2_pbh}
\Delta^2_{\zeta}(k_{\rm PBH})  \sim 10^{-2},
\end{equation}
so that enough regions are over the collapse threshold $\zeta_c \sim 1$ to form PBHs in sufficient abundance to be the dark matter \cite{Motohashi:2017kbs,Passaglia:2018ixg}. 
Moreover, the large-scale modes should be sourced predominantly by a single degree of freedom in order to comply with isocurvature constraints from the CMB. 
The working assumption for a successful model is that inflaton perturbations lead to a
$\zeta$ which dominates on CMB scales while Higgs fluctuations produce one which
dominates on PBH scales. We use a superscript $k$ to denote relations that are exclusively in
Fourier space as opposed to real-space quantities or linear relations that apply to both.

Although the Higgs field $h$ is a spectator during inflation, the mechanism works by enhancing its impact on the total $\zeta$ by exploiting the unstable, unbounded nature of the Higgs potential $V(h)$ at large field values $h>\hcr$ in the Standard Model. 
In particular, the effective Higgs potential at field values far larger than its electroweak vacuum expectation value can be approximated as \cite{Gross:2018ivp,Espinosa:2015qea}
\begin{equation}
V(h) = \frac{1}{4} \lambda h^4,
\end{equation}
with $\lambda=\lambda^{\SM}$ and
\begin{equation} \label{eq:lamsm}
\lambda^{\SM}  \simeq -b \ln \left(\frac{h^2}{\hcr^2 \sqrt{e}}\right),
\end{equation}
where $\hcr$ is the location of the maximum of the Higgs potential which separates the familiar metastable electroweak vacuum from the unstable region, and $b$ controls the flatness of the potential around the maximum. $\hcr$ and $b$ are computable given the parameters of  the SM, and in this work, we choose to fix them at representative values $\hcr = 4 \times 10^{12} \GeV$ and $b = 0.09 / (4 \pi)^2$, corresponding to a top quark mass $M_t \simeq 172 \GeV$, following Refs.~\cite{Espinosa:2017sgp,Espinosa:2018euj,Gross:2018ivp} to facilitate comparisons. 
The Higgs instability exists for $M_t \gtrsim 171 \GeV$ \cite{Buttazzo:2013uya}, which includes
the range from the most recent constraints by the Tevatron and the LHC \cite{ATLAS:2014wva,Khachatryan:2015hba,Aaboud:2016igd,TevatronElectroweakWorkingGroup:2016lid}. Here we have neglected an effective mass term for the Higgs generically generated by a nonminimal Higgs coupling to the Ricci scalar, since at the level expected from quantum corrections it does not change the qualitative features of the mechanism \cite{Gross:2018ivp}. 

\begin{figure}[t]
\includegraphics{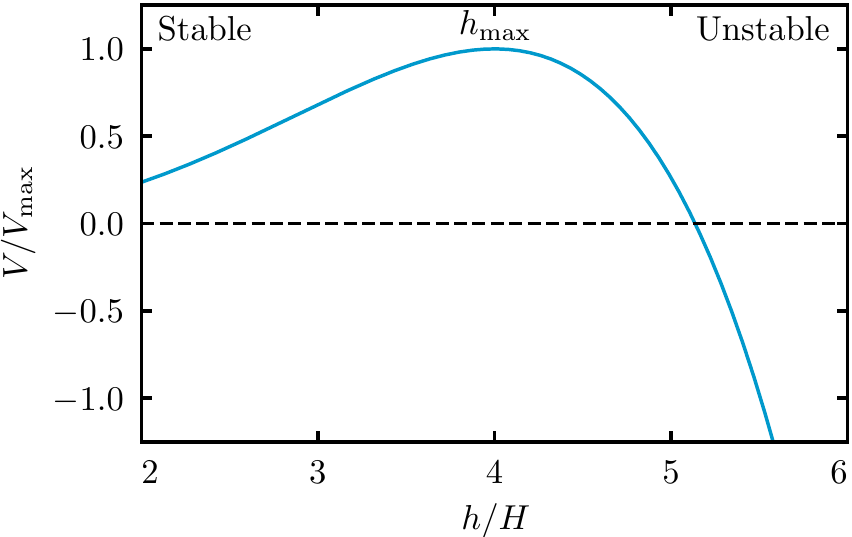}
\caption{Beyond the field value $\hcr$, the Standard Model Higgs effective potential turns over and decreases rapidly, with an unbounded true vacuum of negative energy density. In the PBH scenario, the spectator Higgs rolls down the unstable potential, amplifying its stochastic field fluctuations.
}
\label{fig:potential_turn_over}
\end{figure}

\begin{figure}[t]
\includegraphics{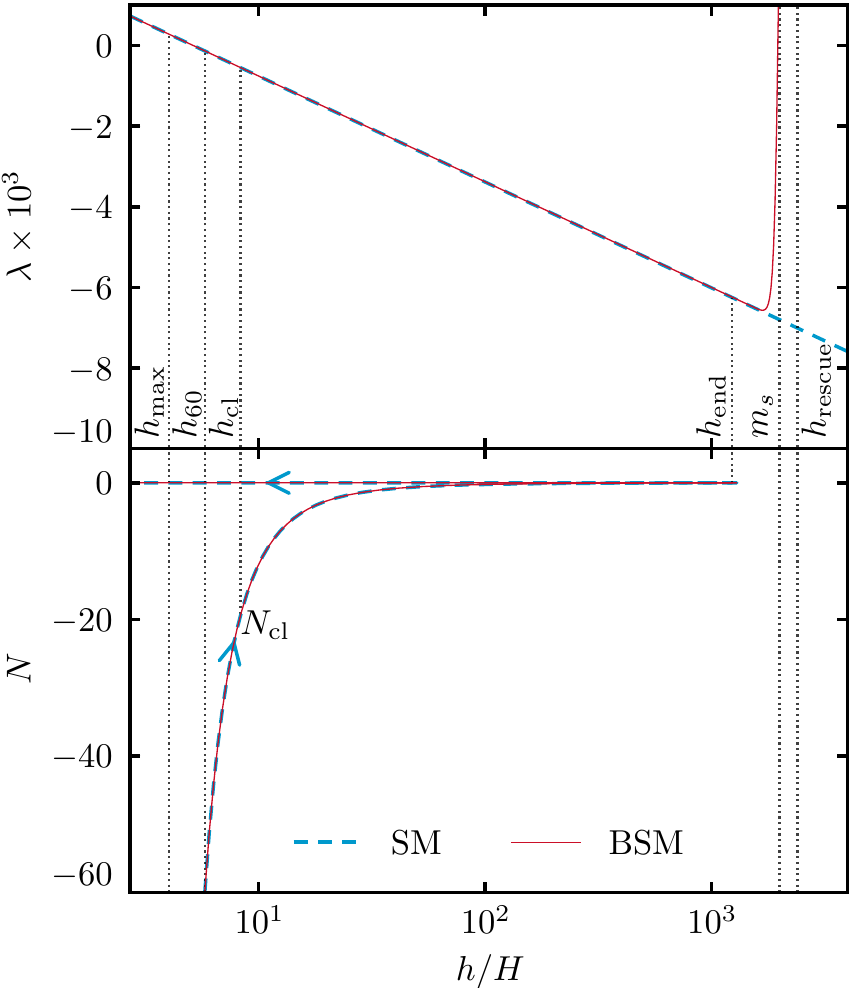}
\caption{Quartic coupling $\lambda$ and $e$-folds $N$ corresponding
to the Higgs field position $h$, with marked special values as computed in \S\ref{sec:inflation} and \S\ref{sec:reheating}: $\hcr$, the maximum of the Higgs potential; $h_{60}$, its position 60 $e$-folds before the end of inflation;  $\hcl$, its classical roll position; $\hend$, its position at the end of inflation; $\hrescue$, beyond which the SM Higgs cannot be ``rescued" by reheating so that it rolls back and oscillates around the origin after inflation. The BSM Higgs adds a coupled scalar of suitable mass $m_s$ to eliminate the runaway instability. 
}
\label{fig:lambdaN}
\end{figure}

We show the potential just around its maximum in Fig.~\ref{fig:potential_turn_over}, and across a wider range of scales in Fig.~\ref{fig:lambdaN} on the unstable side in order to illustrate the
field values that will be crucial to the Higgs instability phenomenology detailed in the following sections. 
Specifically, for  a representative choice for the Hubble scale during inflation $H=10^{12} \GeV$ which we employ throughout for illustration,
the potential maximum is at
\begin{equation}
\label{eq:hmax}
\hcr = 4H.
\end{equation}
When $h<\hcr$, the minimum corresponds to our familiar electroweak vacuum, while for $h>\hcr$, the potential decreases and is unbounded from below. 

On the unstable side of the potential, the Higgs at first rolls slowly relative to the Hubble rate before accelerating as it rolls down the instability.  The location where the Higgs' roll in one $e$-fold becomes comparable to $H/2\pi$ defines the classical roll scale $\hcl$, which we shall define precisely in \S\ref{sec:inflation}. For our parameter choices, it lies at
\begin{equation}
\label{eq:hcl}
\hcl \simeq 8.3 H.
\end{equation}
At the scale $k_{\rm cl}$ which crosses the horizon at $N_{\rm cl} \equiv N(\hcl)$, the power spectrum of $\zeta_h$ at horizon crossing becomes order unity,
\begin{equation}
\label{eq:delta_2_h_rough}
\Delta^2_{\zeta_h} (k_{\rm cl}) \sim 1.
\end{equation}
Our working assumption is that the classical-roll scale $k_{\rm cl}$ will be the one to produce primordial black holes, $k_{\rm cl} = k_{\rm PBH}$. We will show in \S\ref{sec:inflation} that this implies that the Higgs is in fact on the unstable side of the potential during all phases of inflation relevant for observation. In particular, the CMB scales left the horizon during inflation a few $e$-folds after
our Hubble patch crossed the horizon, which we will assume is 60 $e$-folds before the end of inflation. 
At this time, the Higgs is on the unstable side of the potential at
\begin{equation}
\label{eq:h60}
h_{60} \simeq 5.8 H,
\end{equation}
if we take the field value at the end of inflation to be
\begin{equation}
\label{eq:hend}
\hend \simeq 1200 H,
\end{equation}
which we will see below is approximately the largest value possible.  This also leads to $N_{\rm cl} \sim -20$.
If PBHs are produced on that scale then they  have a small mass $M_{\rm PBH} \simeq 10^{-15} M_{\odot}$ at formation, and after mergers and accretion could today lie in the region $M_{\rm PBH} \simeq 10^{-12} M_{\odot}$ where 
all the dark matter could be in the form of PBHs \cite{Montero-Camacho:2019jte,Niikura:2017zjd}.

The Higgs continues to roll to larger field values until the end of reheating when interactions with the thermal bath lift the effective Higgs potential. If the Higgs lies within a maximum rescuable distance $\hrescue$,
\begin{equation}
\hrescue \simeq 2400 H,
\end{equation}
which we compute in \S\ref{sec:reheating}, then after inflation it rolls safely back to the metastable electroweak vacuum, oscillating in a roughly quadratic potential with a temperature-dependent mass $\propto T \propto 1/a$ until it decays to radiation on a uniform Hubble surface. 

The behavior of Higgs perturbations through reheating is complicated. However, all relevant physical scales are at this stage far outside the horizon. So long as a gradient expansion holds,
under which such perturbations can be absorbed into an approximate FLRW background for a local observer, then the curvature field after horizon crossing evolves locally, with no explicit scale dependence \cite{Wands:2000dp}. Therefore the Higgs contribution to the fully nonlinear curvature field $\zeta$ on a uniform total density slice after the Higgs decays is related to the curvature $\zeta_h$ on constant Higgs slices on superhorizon scales during inflation by 
\begin{equation}
\label{eq:zetaconv}
\left. \zeta \right\vert_{{\rm decay}} = R(\zeta_h) \times \left. \zeta_h \right\vert_{\rm inflation},
\end{equation}
with all of the complicated physics of reheating absorbed into a local remapping $R(\zeta_h)$.
More generally, this local remapping would also involve inflaton curvature fluctuations but
these are statistically independent and can be calculated separately in the usual way.
In linear theory, the mapping becomes a simple rescaling factor
\begin{equation}
\label{eq:Rdef}
R \equiv \lim_{\zeta_h \rightarrow 0} R(\zeta_h).
\end{equation} 
The power spectrum after Higgs decay is then related mode by mode to the power spectrum during inflation in a scale-independent fashion
\begin{equation}
\label{eq:D2conv}
\left. \Delta^2_\zeta (k) \right\vert_{{\rm decay}}  = R^2 \left. \Delta^2_{\zeta_h}(k)\right\vert_{\rm inflation},
\end{equation}
in which the number $R$ encodes all of the details of reheating.

It is therefore important to emphasize here that if the Higgs instability mechanism is successful, such that $\Delta_\zeta^2 \sim 10^{-2}$ on PBH scales after the Higgs decays, then in linear theory $R$ must reach at least $0.1$ for the order unity Higgs perturbations~\eqref{eq:delta_2_h_rough} to be converted to curvature perturbations with the correct amplitude~\eqref{eq:delta_2_pbh} to form sufficient PBHs and the details of how this is achieved through reheating and Higgs decay are irrelevant for the prediction of the linear power spectrum on other scales. In particular, its value on CMB scales depends solely on the inflationary $\Delta_{\zeta_h}^2(k)$.
This form is controlled by the Higgs potential itself and by the evolution of the Hubble rate during inflation.
Therefore, the viability with respect to CMB anisotropies of the PBH formation scenarios introduced in the literature \cite{Espinosa:2017sgp,Gross:2018ivp,Espinosa:2018euj}, which all assume mode evolution can be calculated linearly through reheating, can be assessed independently of the
details of the reheating model.

On the other hand, we shall see that the nonlinear nature of the Higgs instability plays
an important role in the mapping between $\zeta_h$ and $\zeta$ in Eq.~\eqref{eq:zetaconv}. 
Here, though the mapping remains local in that a given value of $\zeta_h(\vec{x})$ at a given position
$\vec{x}$ is mapped onto 
a specific value of $\zeta(\vec{x})$, Fourier modes no longer evolve independently.   
Instead, we will compute the mapping using the nonlinear $\delta N$ formalism.  This mapping does depend on the specifics of how inflation ends,  but is independent of physical 
scale.  CMB scale fluctuations are in principle calculable from the spatial field  $\zeta_h(\vec{x})$ determined by modes that froze out during inflation. 

To illustrate these concepts, we make a few simplifying assumptions about how inflation and reheating proceed.  
We show in \S\ref{sec:inflation} that the most optimistic case for the scenario occurs when $H$ is effectively constant through inflation (see Eq.~\eqref{eq:zeta_k_h_comparison}).  Therefore rather than introducing a specific inflaton potential we assume that inflation occurs at a fixed $H$ and ends after an appropriate number of $e$-foldings.
The constant Hubble scale during inflation $H$
and the position $\hend$ 
of the Higgs at the end of inflation then together control the number of $e$-folds between the classical-roll scale $\hcl$ and the end of inflation, and therefore they control the physical scales on which PBHs are formed. 

We then assume that at the end of inflation, the inflaton decays instantly into radiation
and that the Higgs later also suddenly decays into radiation, as in the  model proposed in Ref.~\cite{Espinosa:2017sgp}. Maximizing $R$ in linear theory requires that the position of the Higgs at the end of inflation, $\hend$, is as close as possible to the maximum rescuable distance $\hrescue$. This {criticality} requirement motivates the various choices of scale in Eqs.~\eqref{eq:hmax}-\eqref{eq:hend}, following Ref.~\cite{Espinosa:2017sgp}.  Once $\hend$ is set in this way,
the value $H = 10^{12} \GeV$ is chosen to give a certain mass scale to PBHs by fixing
$N_{\rm cl} \sim -20$.

However, evolving the Higgs on the unstable side of its potential during inflation is dangerous, and the required proximity of $\hend$ to $\hrescue$ aggravates the situation beyond linear theory.
Due to quantum fluctuations of the Higgs during inflation, there are regions in which the local Higgs value at the end of inflation exceeds the background value, overshoots $\hrescue$, and cannot be restored by reheating to the metastable electroweak vacuum created thermally. Such vacuum decay bubbles, with infinitely growing $|\rho_h|$, expand even after the end of inflation and eventually engulf our current horizon.  These quantum fluctuations occur independently in the $e^{120}$ causally disconnected regions at ${N_{\rm cl}}\sim-20$ which make up our current horizon,  and therefore avoiding the vacuum decay bubbles requires extreme fine-tuning
\cite{Gross:2018ivp}. In \S\ref{ssec:nonlin}, we will cast this fine-tuning in terms of a breakdown in linear theory at the end of inflation, and we will show using the nonlinear $\delta N$ formalism that fine-tuning away the vacuum decay bubbles directly tunes away the PBH abundance.

Vacuum decay bubbles can be avoided by stabilizing the Higgs at some large field value between $\hend$ and $\hrescue$.
By adding a singlet heavy scalar of mass $m_s$ with appropriate couplings to the theory, a threshold effect can be exploited to lift the Higgs effective potential during inflation and induce a new true minimum at $h \sim m_s$, preventing unbounded runaway~\cite{EliasMiro:2012ay,Espinosa:2018euj}. For the purposes of this mechanism this Higgs potential beyond the Standard Model can be modeled as
\begin{equation}
\lambda^{\BSM} \simeq \lambda^{\SM} + \frac{\delta \lambda}{2} \left[ 1 + \tanh\left( \frac{h - m_s}{\delta} \right) \right],
\end{equation}
such that for $h \ll m_s$ the potential is as in the SM, given in Eq.~\eqref{eq:lamsm}, while for $h\gg m_s$ the potential is increased by $\delta \lambda ( h^4/4)$. The step height $\delta \lambda$ should be such that the Higgs potential is stabilized, the step position $m_s$ should be close to $\hrescue$, and the step width $\delta$ sufficiently narrow to not interfere with $\hend$. In Fig.~\ref{fig:lambdaN}, we plot this potential with the representative choices $\left\{ \delta \lambda, m_s, \delta \right\} = \left\{0.02,\ 2000 H ,\ 100 H \right\}$.  
While the BSM potential does not suffer from vacuum decay bubbles, it still experiences a breakdown in linearity at the end of inflation. We will therefore also use the nonlinear $\delta N$ formalism to compute the conversion of $\zeta_h$ to $\zeta$ in this case.

Despite this difference at the end of inflation, the SM and BSM potentials are identical until large field values and therefore $\Delta_{\zeta_h}^2$ during inflation is the same in both potentials.
Fluctuations at this stage can be locally
remapped onto $\Delta_{\zeta}^2$.  
For a successful PBH model, this remapping must still achieve
$\Delta_{\zeta}^2 \sim 10^{-2}$ in both cases.
We will therefore focus on the SM potential until we begin discussing nonlinear effects at the end of inflation in \S\ref{sec:reheating}.

\section{Inflationary Higgs Spectrum}
\label{sec:inflation}

In this section, we compute the power spectrum of the Higgs fluctuations from their production inside the horizon through to a common epoch when all modes relevant for observation are superhorizon in scale.

In \S\ref{ssec:classicalroll}, we present the equation of motion for the Higgs and describe the local competition between stochastic kicks and classical roll which governs its evolution. In \S\ref{ssec:background}, we argue that a well-defined background for the Higgs exists during the inflationary epochs relevant for observations, and that Higgs fluctuations during inflation can be computed by linearizing around this background mode by mode. In \S\ref{ssec:attini}, we follow this procedure and compute the Higgs power spectrum during inflation at all scales relevant for observations, 
regarding the Higgs field as a spectator and hence dropping metric perturbations.
In App.~\ref{app:nonini}, we show that our results are consistent with the creation of our background by superhorizon stochastic kicks.
In App.~\ref{app:nonadiabatic} we discuss the role of metric perturbations and nonadiabatic pressure in the evolution of Higgs fluctuations.

Combined, these arguments will show that during inflation the Higgs power spectrum at CMB scales is larger than the Higgs power spectrum at the primordial black hole scales
\begin{equation}
\Delta^2_{\zeta_h}(k_{\rm CMB})  > \Delta^2_{\zeta_h}(k_{\rm PBH}).
\end{equation}

After the conversion of these superhorizon Higgs fluctuations during inflation into curvature fluctuations after inflation through Eq.~\eqref{eq:zetaconv}, this leads to
\begin{equation}
\Delta^2_{\zeta}(k_{\rm CMB})  > \Delta^2_{\zeta}(k_{\rm PBH}),
\end{equation}
in linear theory. 
Accounting for nonlinearity, we shall see that
a similar relation between the scales holds as long as the mapping between the Higgs and curvature fluctuations
is local.
Therefore the first conclusion of the present paper is that in the Higgs vacuum instability scenario, a large amplitude of the power spectrum on small scales generating PBHs is ruled out by the CMB normalization.
Conversely,
if one chooses a different set of parameters in this scenario in order to satisfy the CMB normalization, one ends up with a small-scale power spectrum of at most $\O(10^{-9})$ which fails to form PBHs. 

\subsection{Classical Roll vs Stochastic Kicks}
\label{ssec:classicalroll}

The equation of motion for the position- and time-dependent Higgs field $h(\vec{x}, N)$ is the Klein-Gordon equation
\begin{equation}
\Box h(\vec{x},N) = \frac{\pa V}{\pa h}\bigg\vert_{h(\vec{x},N)} \equiv \left. V_{,h}\right\vert_{h(\vec{x},N)} ,
\label{eq:KG}
\end{equation}
where here and throughout we denote partial derivatives with comma subscripts for compactness.

An important scale in this equation is the classical-roll scale $\hcl$, defined as follows. Every $e$-fold, the potential derivative leads $h(\vec{x},N)$ to roll by
\begin{equation}
\Delta h \simeq -\frac{1}{3 H^2} \left.V_{,h}\right\vert_{h(\vec{x},N)} .
\end{equation}
Meanwhile, if one splits the field into a piece averaged on scales larger than a fixed proper distance $\sim 1/H$ and small scale modes which continually cross the averaging scale, the small scale modes can be viewed as providing a local stochastic noise term to the equation for the coarse-grained superhorizon field~\cite{Starobinsky:1986fx,Starobinsky:1994bd}. The rms of this noise term each $e$-fold is
\begin{equation}
\langle \Delta h ^2 \rangle^{\frac{1}{2}} \simeq \frac{H}{2 \pi}.
\end{equation}
In the language of perturbation theory, this is the per $e$-fold rms of the free field fluctuation $\delta h$ and leads to a stochastic behavior of $h(\vec{x}, N)$. There are no subtleties involved in using $N$ as a time coordinate since the number of $e$-folds is not a stochastic quantity so long as the Higgs remains a spectator. 

The location $\hcl$ in the potential where the roll contribution and the stochastic contribution are equal,
\begin{equation}
\label{eq:classical_roll}
-\frac{1}{3 H^2} \left. V_{,h}\right\vert_{\hcl}   = \frac{H}{2 \pi},
\end{equation}
defines the `classical roll' scale $\hcl$ beyond which the classical term dominates the evolution of $h(\vec{x},N)$. We show this scale in Fig.~\ref{fig:lambdaN}, where it lies at $\hcl \simeq 8.3 H$. 

The classical-roll scale is important because in slow roll, Higgs modes which cross the horizon when the background satisfies Eq.~\eqref{eq:classical_roll} generically have a large power spectrum. In particular, the Higgs power spectrum at the scale $k_{\rm cl}$ which crosses the horizon at $N_{\rm cl}=N(\hcl)$ is order one at horizon crossing,
\begin{equation}
\Delta^2_{\zeta_h} (k_{\rm cl}) \sim \frac{\langle \Delta h ^2 \rangle}{(\Delta h)^2} \sim 1.
\end{equation} 
If these Higgs fluctuations are converted to large curvature fluctuations, they can satisfy the requirements of \S\ref{sec:mechanism} such that $k_{\rm cl} = k_{\rm PBH}$. 

\subsection{Background and Linearization}
\label{ssec:background}

We split $h(\vec{x},N)$ equation into a background and perturbations
\begin{equation}
h(\vec{x},N) = h(N) + \delta h(\vec{x},N).
\end{equation}
Here we define the background to be the part of the field representing
 the spatial average over our  Hubble patch.  Therefore  at 
$N\sim-60$, the spatial average for the perturbation vanishes, $\langle \delta h(\vec{x}, -60 )\rangle = 0$.  The fluctuations are then generated by kicks from quantum fluctuations at
$N > -60$.

To evolve the Higgs field under Eq.~\eqref{eq:KG} we need to evaluate its position on the potential after $N=-60$, as established by its classical roll or quantum kicks. To do this, we need to establish whether the perturbations $\delta h(\vec{x},N)$ are linear around the background $h(N)$.

For the mechanism to work, there should be a well-defined classical roll to the Higgs field
at $N_{\rm cl}$, and we 
can linearize the Higgs fluctuations at that epoch as usual \cite{Espinosa:2017sgp,Gross:2018ivp}.
Between $-60 \lesssim N \lesssim N_{\rm cl}$, there is a competition between the local stochastic kicks and the bulk classical roll, and we need to check whether the kicks destabilize the average field in our Hubble patch.  

\begin{figure}[t]
\includegraphics{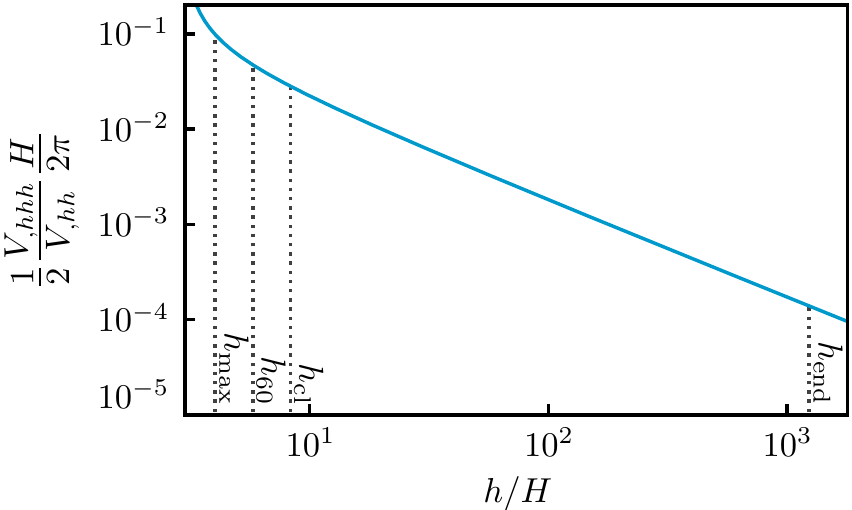}
\caption{Interaction strength for modes of the typical horizon crossing amplitude $H/2\pi$  from \eqref{eq:potential_linearity}. Since it is far less than one at the relevant scales $h_{60}$ and $\hcl$,
modes evolve independently and the cumulative background roll \eqref{eq:background_roll} dominates 
over the stochastic displacement \eqref{eq:stochastic_displacement}.
Linear perturbation theory holds until $\delta h$ grows much larger than $H/2\pi$, near $h_{\rm end}$
(see \S\ref{ssec:nonlin}). }
\label{fig:potential_linearity}
\end{figure}

In linear theory,
stochastic kicks at the same $\vec{x}$ but subsequent times evolve independently from each other. In particular, the potential term in the Klein-Gordon equation controls their interactions. When the Higgs is a spectator field, and we can expand this term around the homogeneous piece as
\begin{equation}
\left.V_{,h}\right\vert_{h(\vec{x},N)} = \left.V_{,h}\right\vert_{h(N)} + \delta h(\vec{x},N) \left. V_{,hh} \right\vert_{h(N)} + \ldots
\end{equation}
where `$\ldots$' contains terms higher order in $\delta h(\vec{x},N)$. If we neglect the higher -order terms, then each subsequent kick evolves as a free field and, as previously mentioned, has rms $H / 2 \pi$ at horizon crossing.    The higher-order terms then are suppressed relative to the linear term by
\begin{equation}
\label{eq:potential_linearity}
\frac{1}{2} \frac{V_{,hhh}}{V_{,hh}} \sqrt{\langle \delta h^2 \rangle}  \simeq \frac{1}{2} \frac{V_{,hhh}}{V_{,hh}} \frac{H}{2 \pi},
\end{equation}
in which we approximate all modes by their value at horizon crossing, which is appropriate while both the inflaton and the Higgs fields 
are slowly rolling. We plot this quantity for the Standard Model Higgs potential in Fig.~\ref{fig:potential_linearity} and show that it is less than one at all scales in the unstable region. 

Therefore, the distance traveled due to stochastic kicks between $N=-60$ and $N_{\rm cl}$ accumulates as a random walk.  For $N_{\rm cl}=-20$,
they therefore lead to a displacement
\begin{equation}
\label{eq:stochastic_displacement}
|\Delta h|_{\rm stochastic} = \frac{H}{2\pi} \times \sqrt{40} \simeq H.
\end{equation}
This is significantly less than the distance between the classical-roll scale $\hcl=8.3 H$ and the maximum of the potential $\hcr=4 H$. This means that stochastic kicks do not, over $40$ $e$-folds, kick our horizon into the other side of the potential. Therefore our whole Hubble volume was on the unstable side of the potential when it crossed the inflationary horizon at $N=-60$.

Moreover, the total displacement from stochastic kicks~\eqref{eq:stochastic_displacement} is less than the amount the background field rolls in these $40$ $e$-folds as we shall now see.
For a homogeneous field $h(N)$, Eq.~\eqref{eq:KG} becomes
\begin{equation}
\label{eq:higgs_bg_eom}
h'' + (3 - \e) h' + \frac{V_{,h}}{H^2} = 0 ,
\end{equation}
where $\e= -H'/H$ is the first Hubble slow-roll parameter, which is zero during the exact de Sitter inflation in our fiducial model. 

The initial conditions $h_{60}$ and $h'_{60}$ for this equation should be such that the Higgs reaches the desired field position $\hend$ at the end of inflation close enough to $\hrescue$ such that
$R$ is maximized. With one constraint and two initial values, a range of  $h'_{60}$ and $h_{60}$ can lead to $h(N=0) = \hend$. Assuming attractor initial conditions for the Higgs, we choose
\begin{equation}
\label{eq:attractorICs}
h'_{60} = -\frac{1}{(3-\e) H^2} \left.V_{,h}\right\vert_{h_{60}},
\end{equation}
making the initial field position given by Eq.~\eqref{eq:h60} $h_{60} \simeq 5.8 H$, when $\hend$ is set by Eq.~\eqref{eq:hend}.

Therefore the classical roll from $-60$ to $N_{\rm cl}$ is 
\begin{equation}
\label{eq:background_roll}
\left\vert\Delta h \right\vert_{\rm roll} = \left\vert h_{60}-\hcl \right\vert = 2.5 H,
\end{equation}
which is significantly larger than the stochastic displacement \eqref{eq:stochastic_displacement} but is nonetheless safely on the unstable side of the potential. This occurs despite the fact that each stochastic kick is larger than the per $e$-fold roll because the roll is coherent across our Hubble volume while the kicks are random.

Therefore we have a consistent picture where if we begin with an average field in our horizon volume around $h_{60} \sim 5.8 H$, then our local background will reach $\hcl$ at $N\sim-20$, unspoiled by stochastic kicks. Between these scales perturbations are linear, thanks to Fig.~\ref{fig:potential_linearity}, and subdominant over the background roll. Our background will then continue to roll to $\hend$, where the Higgs will be uplifted. We plot this background in the lower panel of Fig.~\ref{fig:lambdaN}. In App.~\ref{app:nonini} we show that this picture is consistent with the creation of our background from superhorizon stochastic fluctuations.

We can now use this background to solve for $\delta h$ mode by mode during inflation for all relevant observational scales as in linear theory. This linearization depends on ignoring the interaction of Higgs fluctuations rather than the full machinery of linear perturbation theory for the metric and the matter, and in particular its validity does not assume $|\zeta_h | \ll 1$. Higgs nonlinearities become important in the last $e$-fold of inflation and beyond as the field fluctuations are amplified by the Higgs instability. Such nonlinear effects will affect the superhorizon CMB and PBH modes equally as we shall show in  \S\ref{sec:reheating}.

\subsection{Higgs Power Spectrum}
\label{ssec:attini}

The linearized Klein-Gordon equation for the Fourier mode $\delta h^k(N)$ of $\delta h(\vec{x},N)$ in spatially flat gauge is
\begin{equation}
\label{eq:linKG}
\left(\frac{d^2}{d \eta^2} + 2 \frac{\dot a}{a} \frac{d}{d \eta} + k^2   \right) \delta h^k + a^2  \delta V^k_{,h} \simeq 0 ,
\end{equation}
where $\delta V^k_{,h} = V_{,hh} \delta h^k$ during inflation.  Here we have dropped metric perturbations, which are suppressed when the Higgs is a spectator; we restore these in App.~\ref{app:nonadiabatic} for completeness (see \eqref{eq:linKGwithmetric}).

The Klein-Gordon equation can then be conveniently expressed in terms of the auxiliary variable $u^k \equiv a \delta h^k$,  
\begin{align*}
\ddot{u}^k + \left(k^2 - \frac{\ddot{a}}{a} + a^2 V_{,hh} \right) u^k = 0,
\numberthis
\end{align*}
which holds at all orders of background and Higgs slow-roll parameters. 
To order $\O(\e^2)$ and $\O(\e \eta_H)$, where $\eta_H$ is the second Hubble roll parameter (see, e.g., Ref.~\cite{Miranda:2012rm}), but fully general in terms of the Higgs roll, we can write
\begin{equation}
\label{eq:MukhanovSasaki}
\ddot{u}^k + \left(k^2 - \frac{\ddot{z}}{z}\right) u^k = 0,
\end{equation}
where 
\begin{equation}
z \equiv H \dot{h}.
\end{equation}
This equation, of the Mukhanov-Sasaki type, is conveniently solved in the variable $s \equiv
\eta_{\rm end}-\eta$, the positive decreasing conformal time to the end of inflation (see, e.g., Refs.~\cite{Motohashi:2015hpa,Motohashi:2017gqb}). 

First, let us focus on the evolution in the superhorizon regime.
In that limit, the analytic solution for the Mukhanov-Sasaki equation~\eqref{eq:MukhanovSasaki} is given by
\begin{equation}
\frac{u^k}{z} = c_0 + c_1 \int \frac{d \eta}{z^2},
\end{equation}
where $c_0$ and $c_1$ are constants. So long as the second mode is decaying, we therefore have that on superhorizon scales
\begin{equation}
\label{eq:ms_uoverz}
\frac{\delta h^k}{h'} = c_0 H^2 + \O(\e, \eta_H).
\end{equation}
In linear theory, the curvature perturbation on uniform Higgs density slices is obtained by gauge transformation as
\begin{equation}
\label{eq:infinitesimal}
\zeta^k_h = -\frac{\delta \rho^k_h}{\rho_h'} \simeq -\frac{\delta h^k}{h'},
\end{equation}
where first the approximate equality indicates that when the Higgs is slowly rolling, uniform Higgs density and uniform Higgs field slicing coincide to order $\e$ (see App.~\ref{app:nonadiabatic}). More generally $\zeta_h$ is defined as the change in $e$-folds from a spatially flat surface to a constant density Higgs surface.  This linear approximation holds so long as $\rho_h''/\rho_h'^2 \delta \rho_h \ll 1$, as it is here (see Eq.~\eqref{eq:delta_N_linear}).

Using the superhorizon evolution equation \eqref{eq:ms_uoverz}, we therefore have that if $H$ evolves during inflation, the curvature on uniform Higgs density slices is not conserved on superhorizon scales and in particular decays according to
\begin{equation}\label{eq:sup}
\frac{{\zeta_h^k}'}{\zeta_h^k} = -2\e,
\end{equation}
at leading order in $\e$. This estimate of superhorizon evolution assumes only background slow roll.

This superhorizon evolution is due to a pressure perturbation on uniform density slices for the Higgs, in other words a nonadiabatic pressure, induced because the uniform Higgs density slicing is not a uniform Hubble slicing when $\e\neq0$.  We study this phenomenon in detail in App.~\ref{app:nonadiabatic}.  Conversely, if $H$ is constant, then the fact that the Higgs field evolves onto an attractor solution implies that nonadiabatic stress vanishes thereafter
and $\zeta_h$ is conserved nonlinearly.   In this case, much like single-field slow-roll inflation, 
the Higgs field supplies the only clock and field perturbations are equivalent to changing 
that clock on the background trajectory. 

Next, let us focus on the evolution from subhorizon scales to the superhorizon regime.
This evolution can be tracked by solving the Mukhanov-Sasaki equation~\eqref{eq:MukhanovSasaki} with Bunch-Davies initial conditions deep inside the horizon 
\begin{equation}
\label{eq:bunch-davies}
u^k ( s ) = \frac{1}{\sqrt{2 k}} \left( 1 + \frac{i}{k s} \right) e^{i k s}.
\end{equation}
For analytic estimates, we can assume slow-roll evolution of $z$, in which case we can take the de Sitter mode function \eqref{eq:bunch-davies} to the superhorizon limit, and find that each field fluctuation crosses the horizon with amplitude
\begin{equation}
\label{eq:hkds}
\delta h^k \simeq \frac{i H}{\sqrt{2 k^3}},
\end{equation}
and the field fluctuation power spectrum at that time is
\begin{equation}
\Delta^2_{\delta h} (k) = \frac{k^3}{2 \pi^2} \left\vert \delta h^k \right\vert^2 \simeq \left(\frac{H}{2 \pi}\right)^2 . 
\end{equation}
Using the gauge transformation Eq.~\eqref{eq:infinitesimal} with the field fluctuation Eq.~\eqref{eq:hkds} and the field velocity from the slow-roll solution of Eq.~\eqref{eq:higgs_bg_eom}, the curvature perturbation on uniform Higgs density hypersurfaces at horizon crossing is 
\begin{align*}
\label{eq:zeta_k_h_HC}
\zeta^k_h (\eta_k) &\simeq -\frac{i H}{\sqrt{2 k^3}} \frac{1}{h'} \bigg\vert_{\eta_k}\\
&\simeq -\frac{i H}{\sqrt{2 k^3}}\frac{(3-\e) H^2}{-V_{,h}} \bigg\vert_{\eta_k} . \numberthis
\end{align*}
To lowest order in Higgs- and background-slow-roll $\eta_k$
is chosen to be the epoch of horizon crossing $k \eta_k = 1$, but to next order
can be optimized to $k \eta_k = \exp{\left[7/3 - \ln 2 - \gamma_E\right]}$, with $\gamma_E$ the Euler-Mascheroni constant  \cite{Motohashi:2015hpa,Motohashi:2017gqb}.

We can now estimate the relative amplitude of $\Delta^2_{\zeta_h}$ on CMB and PBH scales. Choosing some comparison time $\eta_*$ once both scales have exited the horizon but far enough from the end of inflation that slow-roll parameters are still small, we have 
\begin{align*}
\label{eq:zeta_k_h_comparison}
\left.\frac{\Delta^2_{\zeta_h} (k_{\rm CMB})}{\Delta^2_{\zeta_h} (k_{\rm PBH})}\right\vert_{\eta_*}  &\simeq \left( \frac{H_{\rm PBH}}{H_{\rm CMB}}\right)^4 
\frac{k_{\rm CMB}^{3} \left\vert\zeta^{k_{\rm CMB}}_h\right\vert^2_{\eta_k}}{k_{\rm PBH}^{3} \left\vert\zeta^{k_{\rm PBH}}_h\right\vert_{\eta_k}^2}\\
&\simeq \left(\frac{H_{\rm CMB}}{H_{\rm PBH}}\right)^2 \left(\frac{V_{,h} |_{\rm PBH}}{V_{,h} |_{\rm CMB}}\right)^2,
\numberthis
\end{align*}
where we have used at horizon crossing the Higgs slow-roll expression \eqref{eq:zeta_k_h_HC} and outside the horizon the Hubble slow-roll expression \eqref{eq:ms_uoverz}. Thus in the generic situation where $H$ is decreasing and the Higgs rolls downhill, we find that $\Delta^2_{\zeta_h} (k_{\rm CMB}) / \Delta^2_{\zeta_h} (k_{\rm PBH}) > 1$. The most optimistic case for the scenario is therefore the one where $H$ is strictly constant between the CMB and PBH scales, and it still results in $\Delta^2_{\zeta_h} (k_{\rm CMB}) / \Delta^2_{\zeta_h} (k_{\rm PBH}) > 1$.

\begin{figure}[t]
\includegraphics{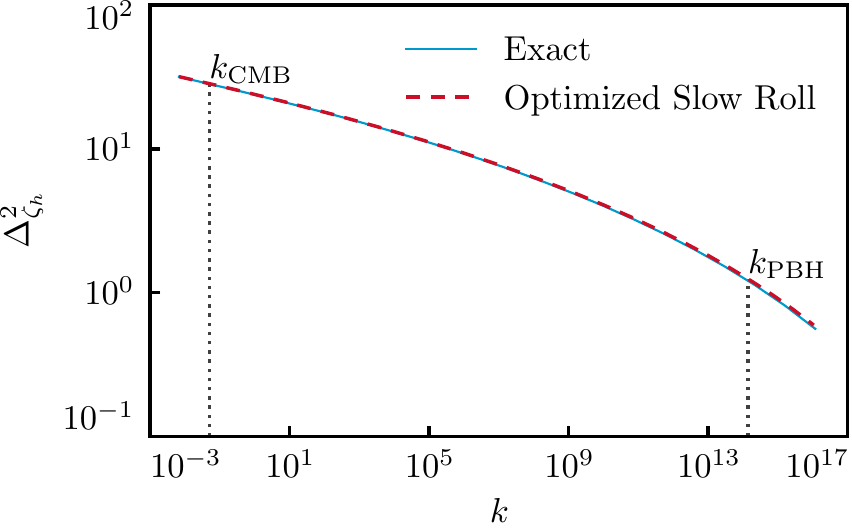}
\caption{
The  Higgs power spectrum during inflation computed as described in \S\ref{ssec:attini} on a uniform Higgs energy density slice by exact solution of the Mukhanov-Sasaki equation \eqref{eq:MukhanovSasaki} (solid blue) and by the optimized slow-roll approximation \eqref{eq:zeta_k_h_HC} (dashed red), at some time $\eta_*$ after the relevant modes have crossed the horizon in the optimistic scenario where $H$ is constant until $\eta_*$. The Higgs power spectrum is larger on CMB scales than on the classical-roll scales.
}
\label{fig:higgspowerspec}
\end{figure}
We show in Fig.~\ref{fig:higgspowerspec} the  Higgs power spectrum computed by solving the Mukhanov-Sasaki equation exactly in the optimistic case where $H$ is constant between CMB and classical-roll scales, evaluated at the convenient time $\eta_*$ when all relevant modes have crossed the horizon. We compare this exact solution to the slow-roll expression \eqref{eq:zeta_k_h_HC} with the optimized freeze-out epoch. 
For decreasing $H$, the ratio $\Delta^2_{\zeta_h} (k_{\rm CMB})/\Delta^2_{\zeta_h} (k_{\rm PBH})$ would be larger than the one estimated from Fig.~\ref{fig:higgspowerspec}.

These Higgs fluctuations at $\eta_*$ will be converted to total curvature fluctuations after Higgs decay by the factor $R(\zeta_h)^2$ which we discuss in \S\ref{sec:reheating}.
The key is that this mapping affects all mode contributions to $\zeta_h$ uniformly.  For example in linear theory $R$ is a constant whose value must be $\sim 0.1$ for successful PBH formation at $k_{\rm PBH}$.  Eq.~\eqref{eq:zeta_k_h_comparison} determines the total curvature power relative to this scale.  In particular, the power spectrum at CMB scales is an order of magnitude larger than the power spectrum at the classical-roll scale. It is simply a feature of the Higgs potential that the field slope increases as the Higgs goes farther into the unstable region, and therefore that the Higgs fluctuation shrinks as $k$ increases. Thus, a model with $\Delta^2_{\zeta_h} (k_{\rm CMB}) / \Delta^2_{\zeta_h} (k_{\rm PBH}) > 1$ that forms
PBHs at $k_{\rm PBH}$ will necessarily violate CMB constraints.

The results of this section hold equally for the SM and BSM potentials. The difference between the two potentials enters only into $R(\zeta_h)$ which converts these results into the final curvature perturbation after Higgs decay. More generally as long as this mapping depends only on field amplitude and not on $k$ explicitly, PBHs cannot be formed from
{the} Higgs instability without violating CMB constraints.

\begin{center}
  $\ast$~$\ast$~$\ast$
\end{center}

In summary, we have shown that during inflation
\begin{equation}
\Delta^2_{\zeta_h}(k_{\rm CMB}) > \Delta^2_{\zeta_h}(k_{\rm PBH}).
\end{equation}
As we argued in \S\ref{sec:mechanism}, the conversion of the inflationary $\zeta_h$ to the final $\zeta$ depends only on the amplitude of $\zeta_h$ and thus all the information about reheating can be encoded in a scale-independent function $R(\zeta_h)$. 

This means that in linear theory, where $R$ is a constant, if primordial black holes are produced on small scales then on large scales
\begin{equation}
\Delta^2_{\zeta}(k_{\rm CMB})  > \Delta^2_{\zeta}(k_{\rm PBH}) \gtrsim 10^{-2},
\end{equation}
which is incompatible with measurements of the CMB. 

Nonlinearly, when CMB and PBH modes cannot be tracked independently through the final $e$-folding of inflation and reheating, the $\Delta^2_{\zeta_h}(k)$ results in this section provide the superhorizon initial conditions which can be mapped to the final $\zeta$. Given that this  local mapping $R(\zeta_h)$ does not distinguish
between Higgs fluctuations of  different physical scales, we will argue in \S\ref{sec:reheating} that even nonlinearly, Higgs induced curvature
fluctuations
on CMB scales will be larger than those on PBH scales.

We now study in \S\ref{sec:reheating} the specific values taken by the conversion function $R(\zeta_h)$ itself. This will allow us to determine whether or not the Higgs fluctuations computed here can be transferred into large enough curvature perturbations to form PBHs, regardless of the compatibility with the CMB. Moreover, given that we have produced large Higgs fluctuations on CMB scales, this will allow us to determine under which conditions Higgs 
criticality is incompatible with the small curvature fluctuations observed in the CMB.

\section{Curvature Fluctuations from Reheating}
\label{sec:reheating}

We now track the curvature perturbations $\zeta_h$ on uniform Higgs density slices during inflation, computed in \S\ref{sec:inflation}, through the end of inflation, reheating, and Higgs decay to compute the final curvature perturbations $\zeta$ on uniform total density hypersurfaces relevant for PBH formation.

In \S\ref{ssec:instantaneous}, we discuss how the Higgs evolves near the end of inflation and present the basic features of the instantaneous reheating model proposed by Ref.~\cite{Espinosa:2017sgp}.

In \S\ref{ssec:deltaN}, we discuss how to use the nonlinear $\delta N$ formalism to convert $\zeta_h$ to $\zeta$ for any local reheating scenario, and we discuss jump conditions which much be satisfied during instantaneous reheating. We also present linearized $\delta N$ formulae which yield results corresponding to those of linear perturbation theory, allowing an important cross-check of the computation.

In \S\ref{ssec:linear}, we follow the assumption of Refs.~\cite{Espinosa:2017sgp,Gross:2018ivp,Espinosa:2018euj,Espinosa:2018eve} that linear theory holds through reheating and we compute explicitly the conversion of the inflationary $\zeta_h$ to the final $\zeta$. We show that energy conservation at reheating, neglected in previous works, prevents the model from achieving the required $R=0.1$ for both the SM and BSM potentials and thus PBH 
are not produced in sufficient quantities to be the dark matter in linear theory.

In \S\ref{ssec:nonlin} we show that linear theory is in fact violated at the end of inflation and we explicitly compute the full nonlinear conversion $R(\zeta_h)$ for the SM and BSM Higgs effective potentials. We show that PBHs in sufficient abundance to be the dark matter are never formed, second-order gravitational waves are suppressed, and only for a special class of criticality scenarios can observable perturbations be produced on CMB scales.

\subsection{Instantaneous Reheating}
\label{ssec:instantaneous}

\begin{figure}[t]
\includegraphics{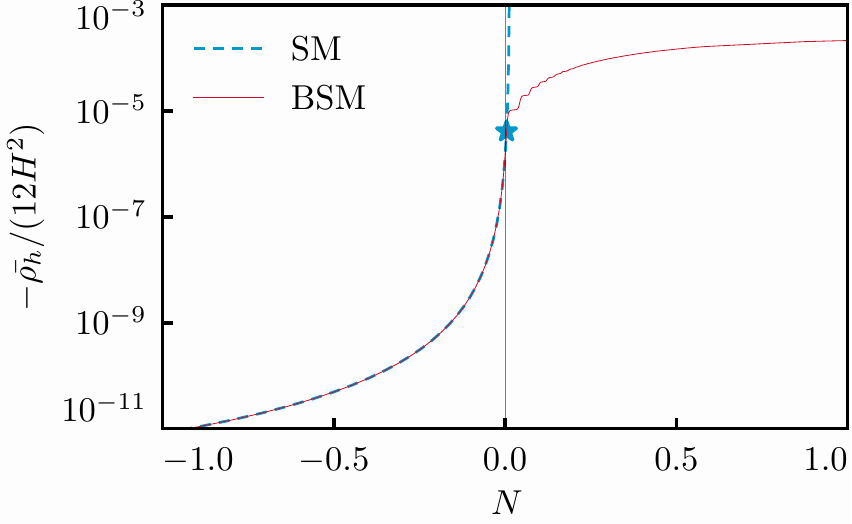}
\caption{The (negative) Higgs energy density evolution during inflation, with $H =\,$const. $N=0$ corresponds to the end of inflation for the background, and $N>0$ shows how the Higgs energy would evolve in
local regions in advance of the background.  
The SM Higgs (dashed blue) at $N=0$ is close to saturating the rescue condition \eqref{eq:rescue} (blue star).
The BSM Higgs (solid red) hits a wall during inflation and therefore is always rescued.
Due to the Higgs attractor behavior, the exact value of $\zeta(0^+)$ for any given local shift $\delta N =\zeta_h$
can be read off using Eq.~\eqref{eq:nonlin_jump}.
}
\label{fig:rho}
\end{figure} 

As the Higgs travels farther and farther on the unstable side of the SM potential, it rolls faster and faster and if inflation never ended its energy density would diverge in finite time. In Fig.~\ref{fig:rho}, we show this $\rho_h$ during this last phase of the instability as a function of the number of $e$-folds.
Of course inflation does end and for our chosen example this occurs at $N=0$  and a field
value  $\hend$ in the background.  In this case, $N>0$ then shows how the Higgs energy would continue to evolve if inflation did not end at $N=0$. This range around $N=0$ will also be useful when we consider perturbations that can
be ahead of or behind the background value.

As we can see, the fiducial position of the background SM Higgs we have chosen is near critical. Its energy is increasing rapidly and if inflation lasts much longer it will gain sufficient energy such that it is no longer a spectator.    On this edge, the {\it background} Higgs field experiences the same evolution in the SM and BSM
cases by construction 
(see Fig.~\ref{fig:rho}).

Once inflation ends, the process of reheating transfers
the inflaton's energy to radiation. The Higgs then no longer evolves in vacuum and the Klein-Gordon equation becomes
\begin{equation}
\Box h = \VT_{,h},
\label{eq:KG_eff}
\end{equation}
where $\VT$ is the thermal effective potential in a bath of temperature $T$ \cite{Espinosa:2015qea},
\begin{equation} \label{eq:VT}
\VT  = V + \frac{1}{2} M_{\rm T}^2 h^2 e^{-h/2 \pi T}, \quad M_{\rm T}^2 \simeq 0.12 T^2.
\end{equation}

If reheating is instantaneous, as proposed by Ref.~\cite{Espinosa:2017sgp}, then the total energy is conserved across it,
\begin{equation}
\left(\rho_{\phi} + \rho_h\right) (0^-) = \left(\rho_{\rm r} + \rho_h\right)(0^+).
\label{eq:energyconback}
\end{equation}

The division of the total energy density after inflation into a radiation piece $\rho_{\rm r}$ and a thermal component to the Higgs  $\rho_h$ is somewhat arbitrary.  Since to leading order in $\rho_h/\rho_{\rm tot}$, the curvature perturbation after the Higgs has decayed depends only on the conservation of the total energy, for convenience we choose to define $\rho_{\rm r}$ as a separately conserved thermal piece obeying the equation of motion
\begin{equation}
\label{eq:rho_r_eom}
\rho_{\rm r}' = - 4 \rho_{\rm r},
\end{equation}
and the thermal bath temperature $T$ is then
\begin{equation}
T = \left(\frac{30 \rho_{\rm r} }{ \pi^2 g_*}\right)^{1/4},
\end{equation}
where $g_*=106.75$ is the number of degrees of freedom in the Standard Model. Conservation of the total stress-energy along with the Higgs equation of motion~\eqref{eq:KG_eff} then imply that the separately conserved Higgs energy is
\begin{equation}
\label{eq:rho_h}
\rho_h \equiv \frac{1}{2} \frac{\dot h^2}{a^2} + \VT.
\end{equation}
Changes in $\VT$ as the Higgs rolls down the effective potential then provide kinetic energy for the field
as if it were a true potential energy.

Since $3 H^2(0^-) = (\rho_\phi + \rho_h){(0^-)}$, neglecting the Higgs' energy density contribution to $H$ during inflation but including it after entails a dynamically negligible $\O(\rho_h/\rho_\phi) \sim \O(\rho_h/3H^2)$  discontinuity
in the Hubble rate at the end of inflation $H_{\rm end}$ (see Fig.~\ref{fig:rho}). On the other hand, strict energy conservation~\eqref{eq:energyconback} at reheating is important because we will evaluate perturbations on constant density surfaces.

After reheating, the thermal effective potential in Eq.~\eqref{eq:KG_eff} will rescue the Higgs from the unbounded SM minimum so long as
\begin{equation}
\label{eq:rescue}
{\rm Max}\left[h\right] < \hrescue,
\end{equation}
where ${\rm Max}\left[h\right]$ is the maximum displacement of the Higgs field and $\hrescue$ is the peak of the thermal potential $\VT$.  Due to the nonzero kinetic energy of the Higgs at the end of inflation, ${\rm Max}\left[h\right]$ is larger than the field displacement at the end of inflation $\hend$. We mark the maximal point which saturates this bound with a star in Fig.~\ref{fig:rho}.

Neglecting the exponential term, we find that the peak of the uplifted potential at reheating is
\begin{equation}
\hrescue^{(0)} = \frac{M_T}{\sqrt{|\lambda^{\SM}|}},
\end{equation}
a solution which can be iterated to account for the exponential term, yielding
\begin{equation}
\label{eq:hrescue}
\hrescue^{(1)} = \hrescue^{(0)} e^{-\hrescue^{(0)} / {4 \pi T}} \sim 1.6 \sqrt{H_{\rm end} \Mpl},
\end{equation}
where we have used $|\lambda| \sim 0.007$ and which with $H_{\rm end}=10^{12} \GeV$ evaluates to $ \sim 2500 H$.
The value of the uplifted Higgs potential at this approximate maximum is
\begin{equation}
\label{eq:potential_max}
\VT(\hrescue^{(1)}) \sim 0.02 H_{\rm end}^2 \Mpl^2.
\end{equation}
These scalings are in good agreement with the exact calculation for $\hrescue^{(0)}/4\pi T \ll 1$ and they serve to highlight the dependence of the results with parameter choices. For our fiducial parameter set, the exact calculation yields  $\hrescue \simeq 2400 H_{\rm end}$, $\VT(\hrescue)\simeq 0.02 H_{\rm end}^2 \Mpl^2$.
 
If the Higgs is rescued, then it oscillates in its uplifted temperature-dependent potential, redshifting as radiation on the cycle average up to corrections from the nonquadratic components of its potential, until it decays on the $e$-fold timescale on constant Hubble surfaces.  The rescue point is therefore relevant even for the BSM potential.   In our example shown in Fig.~\ref{fig:rho}, we set the $m_s$ barrier
close to $\hrescue$ to maximize the instability while ensuring that the field returns to the electroweak
vacuum after reheating.

We now describe in \S\ref{ssec:deltaN} how to track perturbations through the end of inflation and this instantaneous reheating epoch.

\subsection{Nonlinear Curvature Evolution}
\label{ssec:deltaN}

The PBH abundance depends on the probability that the local horizon-averaged density field exceeds some collapse threshold $\delta_c$. We approximate this by the Gaussian probability that the curvature on uniform total density slices $\zeta$ lies above some threshold $\zeta_c$. We therefore need to compute $\zeta$ after the Higgs decays. 

Nonlinearly in the Higgs field perturbations, the transformation of the curvature on uniform Higgs density slicing during inflation $\zeta_h$ to the curvature on uniform total density slicing after inflation $\zeta$ can be performed in the $\delta N$ formalism \cite{Starobinsky:1985aa,Salopek:1990jq,Sasaki:1995aw,Lyth:2004gb}, which allows us to evolve superhorizon perturbations by counting the number of $e$-folds of expansion from an initial flat slice at some convenient initial time $N_i$ to a uniform total density slice at a final time $N$,
\begin{equation}
\label{eq:delta_N}
\zeta(N, \vec{x}) = \N(\rho_\alpha(N_i,\vec{x}) ;\  \rho_{\rm tot} (N)) - \medbar\N( \bar{\rho}_\alpha(N_i);\ \rho_{\rm tot} (N)),
\end{equation}
where $\N$ is the local number of $e$-folds of expansion from the initial flat hypersurface at $N_i$ on which any fields $\alpha$
have energy densities $\rho_\alpha(N_i,\vec{x}) = \bar{\rho}_\alpha(N_i) + \delta \rho_\alpha(N_i, \vec{x})$ to a final surface of uniform total density $\rho_{\rm tot} (N)$, and $\medbar\N$ is the corresponding expansion of the unperturbed universe. 
Using the separate universe assumption, $\N$ can be computed in terms of background FLRW equations for a universe with the labeled energy contents. 

If we chose $N_i$ to be some time during inflation when we know the superhorizon density fluctuation $\delta \rho_h$ (or $\zeta_h$) from our computation in \S\ref{sec:inflation}, then since we are considering only perturbations sourced by the Higgs we can make the inflaton density at $N_i$ implicit and keep only the dependence on the initial $\delta \rho_h$. We will perform this fully general calculation in \S\ref{ssec:nonlin}.

To understand these results, it is also useful to have a simple analytic approximation for the impact of reheating
given conditions just before reheating.  In this case we can set the initial time just after reheating at $N_i=0^+$.
The Higgs energy density is a small component of the total energy budget, and therefore to leading order in $\zeta$ we can evaluate the $\delta N$ formula assuming that $\rho_{\rm tot} \propto a^{-4}$  to find 
\begin{align*}
\zeta(0^+, \vec{x}) &\simeq \frac{1}{4} \ln{\left(\frac{\rho_{\rm tot}(0^+,\vec{x})}{\bar{\rho}_{\rm tot}(0^+)}\right)}\\
&\simeq \frac{\rho_{\rm tot} (0^+,\vec{x})-\bar{\rho}_{\rm tot}(0^+)}{12 H_{\rm end}^2}.
\numberthis
\end{align*}
Conservation of energy at reheating \eqref{eq:energyconback} then implies the $N=0$ jump condition
\begin{equation}
\label{eq:nonlin_jump}
\zeta(0^+, \vec{x}) = \frac{\rho_{h} (0^-,\vec{x})-\bar{\rho}_{h}(0^-)}{12 H_{\rm end}^2}.
\end{equation}
Note that this condition applies to nonlinear Higgs density fluctuations $|(\rho_h -\bar\rho_h)/\bar \rho_h| \gg 1$
so long as $|(\rho_h -\bar\rho_h)/\bar \rho_{\rm tot}| \ll 1$.

We can see from Eq.~\eqref{eq:nonlin_jump} that the postinflationary energy partitioning chosen in Eq.~\eqref{eq:energyconback} does not enter into the total curvature just after reheating. Instead, $\zeta(0^+, \vec{x})$ is determined solely by energy conservation.

We can further simplify this condition by noting that to the extent that $H$ is constant during inflation, which we have shown by Eq.~\eqref{eq:zeta_k_h_comparison} is the most optimistic scenario for PBH production,
the shift in $e$-folds to a constant Higgs energy density $\delta N_h =\zeta_h$ is conserved nonlinearly.\footnote{If $H$ evolves, the leading-order effect will be simply to shrink the Higgs $\delta N_h$.}
Therefore $\rho_{h} (0^-,\vec{x}) = \bar{\rho}_{h} (-\delta N_h)$, 
with this Higgs density computed as though inflation did not end at $N=0$ as in Fig.~\ref{fig:rho}.
We can therefore read off $\zeta(0^+, \vec{x})$ for a given $\zeta_h$ from $\bar\rho_h(N)$ as
\begin{equation}
\label{eq:nonlin_jump2}
\zeta(0^+, \vec{x}) = \frac{\bar\rho_{h} ({-}\zeta_h)-\bar{\rho}_{h}(0^-)}{12 H^2} \Big|_{\rm inf},
\end{equation}
where $|_{\rm inf}$ denotes this convention of evaluating the background as if inflation never ends.

Before using these nonlinear formulae \eqref{eq:delta_N} and \eqref{eq:nonlin_jump2} in \S\ref{ssec:nonlin}, we will in \S\ref{ssec:linear} perform the calculation using linear perturbation theory. To validate the linear theory calculations, below we derive linear approximations to the $\delta N$ formulae. 

The full $\delta N$ formula \eqref{eq:delta_N} can be linearized in $\delta \rho_{h_i} \equiv \delta \rho_h(N_i,\vec{x})$ to obtain
\begin{equation}
\label{eq:delta_N_linear}
\zeta \simeq \frac{\pa \N(\rho_{h_i};\  \rho_{\rm tot} )}{\pa \rho_{h_i}}\delta \rho_{h_i} \simeq  -\frac{\pa \N(\rho_{h_i};\  \rho_{\rm tot})}{\pa \rho_{h_i}}\rho'_{h_i} \zeta_{h_i},
\numberthis
\end{equation}
where $\rho_{h_i}$ and the $\rho'_{h_i}$ are Higgs density and its derivative at $N_i$ and $\zeta_{h_i}$ is the Higgs curvature at that time.
Likewise, a linear Taylor expansion of the jump condition \eqref{eq:nonlin_jump2} is given by
\begin{align*}
\label{eq:linear_jump}
\zeta(0^+, \vec{x}) &\simeq -\zeta_h (0^-) \times \frac{\bar\rho_h'(0^-)}{12 H^2}.
\numberthis
\end{align*}
The ratio $-\rho_h' (0^-) /12 H^2$ is the rescaling factor $R(0^+)$ if $H$ is constant through to the end of inflation.

As we shall see below, these linear $\delta N$ formulae provide an important point of contact between the nonlinear $\delta N$ and linear perturbation theory approaches.

\subsection{Linear Conversion}
\label{ssec:linear}

We now follow the assumption of Refs.~\cite{Espinosa:2017sgp,Gross:2018ivp,Espinosa:2018euj,Espinosa:2018eve} that linear theory holds through reheating, and we show that under this assumption primordial black holes cannot be the dark matter.

While Higgs field values $h$ and $\delta h$ and their derivatives $h'$ and ${\delta h}'$ are all continuous through reheating, the Higgs potential and its slope change instantaneously when the Higgs potential is uplifted. Therefore the Higgs energy density \eqref{eq:rho_h}, its derivative
\begin{equation}
{\rho}_h' = -3 \frac{\dot{h}^2}{a^2}  -  \VT_{,T} T,
\end{equation}
and its perturbation
\begin{equation}
\label{eq:drhohafter}
\delta \rho_h = \frac{1}{a^2} \dot{h} \delta \dot{h}  + \VT_{,h} \delta h +  \VT_{,T} \delta T,
\end{equation}
are not continuous with their values 
at $N=0^-$.
$\delta T$ here is any perturbation in the bath temperature correlated with the Higgs, which we shall see is generically induced at reheating. For simplicity, we have omitted here a contribution to the perturbed energy density coming from the metric lapse perturbation, which we restore in App.~\ref{app:nonadiabatic} 
(see \eqref{deltarho}). 
Its relative contribution is negligible.

The jump in $\rho_h'$ 
\begin{equation}
\Delta \left[\rho_h'\right] =  -\VT_{,T} T,
\end{equation}
and in the energy density perturbation
\begin{equation}
\label{eq:higgs_energy_jump}
\Delta \left[\delta \rho_h\right] = \delta h \left( \VT_{,h}  - V_{,h} \right) +\VT_{,T} \delta T,
\end{equation}
imply that the curvature perturbation~\eqref{eq:infinitesimal} on constant Higgs energy density slices is discontinuous at reheating. This instantaneous change in $\zeta_h$ is due to an instantaneous source in the conservation equation from the interaction of the Higgs with the thermal bath. 

However, the instantaneous increase in the Higgs energy density perturbation $\delta \rho_h$ does not come for free. Conservation of energy, which we imposed at the level of the background in Eq.~\eqref{eq:energyconback}, also holds locally. It implies that the increase in the Higgs energy density is counterbalanced by an induced perturbation in the radiation field
\begin{equation}
\label{eq:delta_rho_r}
\delta \rho_{\rm r} (0^+) = 
- \Delta \left[\delta \rho_h\right],
\end{equation}
and therefore that the Higgs and radiation energy densities after uplift are nearly canceling.
In other words, the uplift creates a Higgs-radiation isocurvature fluctuation rather than a 
net curvature fluctuation. To the extent that the Higgs fluctuation then redshifts like radiation, the isocurvature mode does not subsequently contribute to the curvature fluctuation.

The conserved curvature on uniform radiation density slices which corresponds to this induced radiation perturbation is
\begin{equation}
\label{eq:zeta_r}
\zeta_{\rm r} = - \frac{\delta \rho_{\rm r}}{\rho_{\rm r}'} = \frac{\delta T}{T},
\end{equation}
and solving for the radiation perturbation \eqref{eq:delta_rho_r} using the jump in Higgs energy \eqref{eq:higgs_energy_jump} and 
Eq.~\eqref{eq:zeta_r},
we find
\begin{equation}
\label{eq:radcomp}
\delta \rho_{\rm r} (0^+)  = -\delta h \left( \VT_{,h}  - V_{,h} \right) \left(1- \VT_{,T} \frac{T}{\rho_{\rm r}'}\right)^{-1}.
\end{equation}
This induced radiation perturbation comes from the direct interaction of the Higgs with the radiation during the thermal uplift. It is distinct from radiation perturbations corresponding to intrinsic inflaton fluctuations, which are uncorrelated and can be computed separately, or to inflaton perturbations produced by the gravitational influence of the Higgs perturbations during inflation discussed in App.~\ref{app:nonadiabatic}, which are suppressed. 

This radiation perturbation was omitted in Refs.~\cite{Espinosa:2017sgp,Espinosa:2018eve,Espinosa:2018euj,Gross:2018ivp}, though in fact its role in conserving total energy has a large impact on the final curvature perturbation $\zeta$. In particular, on a constant total density surface, the curvature perturbation is given by Eq.~\eqref{eq:zeta_tot_intro}, reproduced here for convenience,
\begin{equation} 
\label{eq:zeta_tot}
\zeta = \left(1-   \frac{\rho_h'}{\rho_{\rm tot}'} \right)  \zeta_{\rm r} + \frac{\rho_h'}{\rho_{\rm tot}'} \zeta_h.
\end{equation}
Immediately after the uplift of the Higgs potential, $\zeta$ therefore satisfies
\begin{align*}
\label{eq:upliftR}
\zeta (0^+) &= \zeta_h (0^-) \times \frac{\rho_h' (0^-)}{\rho'_{\rm tot} (0^+)} \\ 
&\simeq -\zeta_h (0^-) \times \frac{\rho_h' (0^-)}{12 {H_{\rm end}^2}},\numberthis
\end{align*}
by virtue of Eq.~\eqref{eq:delta_rho_r}. This equation is nothing but the linear jump condition \eqref{eq:linear_jump}, this time derived from linear perturbation theory rather than the $\delta N$ formalism.

To evolve $\zeta(N)$ from its value at $\zeta(0^+)$, we must solve for the Higgs perturbations after uplift. 
Again the most important aspect of this calculation is energy conservation.  
Energy conservation guarantees that the cancellation responsible for the suppression in the starting value
$\zeta(0^+)$ is maintained on a timescale short compared to an $e$-fold.

Although during inflation we assumed the Higgs is a spectator, after potential uplift we jointly solve the Higgs background equation  \eqref{eq:higgs_bg_eom}, now with $V\rightarrow \VT$, and the radiation background equation \eqref{eq:rho_r_eom}, 
with the Hubble rate determined by the Friedmann equation. 
For the perturbations we solve the linearized Klein-Gordon equation \eqref{eq:linKG} for the Higgs, now with a perturbed potential
\begin{equation}
\label{eq:perturbed_potential}
\delta V^{Tk}_{,h} = \VT_{,hh} \delta h^k + \VT_{,hT} \delta T^k,
\end{equation}
which accounts for the effect of temperature perturbations.
Again the metric terms are negligible and we exploit here that $\zeta_{\rm r}$ is constant to avoid solving perturbation equations for the radiation component; we shall show that both are good approximations in App.~\ref{app:nonadiabatic}.

To quantify the importance of the induced radiation perturbation \eqref{eq:delta_rho_r}, we chose $\hend$ such that the postinflationary Higgs contribution to the total $\Delta_\zeta^2 (k_{\rm PBH})$ is $\sim (0.1)^2$. This is the calculation performed in the literature which suggests that primordial black holes can be formed at the classical-roll scale. We then add the induced radiation contribution and see how $\Delta_\zeta$ is affected.

\begin{figure}[t]
\includegraphics{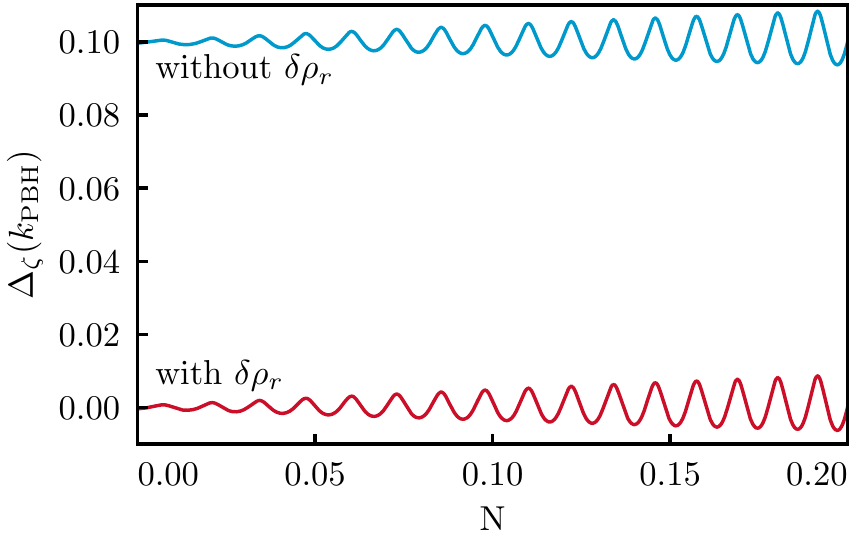}
\caption{The curvature perturbation on uniform total energy 
slices computed in linear perturbation theory 
with and without the inclusion of radiation perturbations required by energy conservation at
reheating.  These induced perturbations suppress the cycle-averaged curvature by several orders of magnitude (see \S\ref{ssec:linear}).}
\label{fig:zeta_comparison}
\end{figure}

We show these numerical results in Fig.~\ref{fig:zeta_comparison}.  
Here we plot $\Delta_\zeta(k_{\rm PBH})$ with a phase convention $\varphi$ such that the analogous superhorizon Higgs fluctuation 
\begin{equation}
\Delta_{\zeta_h} \equiv e^{i\varphi} \sqrt{ \Delta_{\zeta_h}^2},
\end{equation}
is negative real  during inflation. Note that $\Delta_{\zeta_h}$ changes sign at the potential uplift and becomes positive real.
After inflation, $\Delta_\zeta(k_{\rm PBH})$ oscillates between negative and positive values but stays real.

It is immediately clear from Fig.~\ref{fig:zeta_comparison} that the induced radiation perturbation $\delta \rho_{\rm r}$ suppresses the amplitude of the total curvature $\zeta$ by orders of magnitude, making it much more difficult to achieve the required $R = 0.1$ in this model. For the fiducial background, which was claimed to produce $R=0.1$, by taking into account $\delta \rho_{\rm r}$ we instead have $R(0^+)\simeq 3\times10^{-4}$.

\begin{figure}[t]
\includegraphics{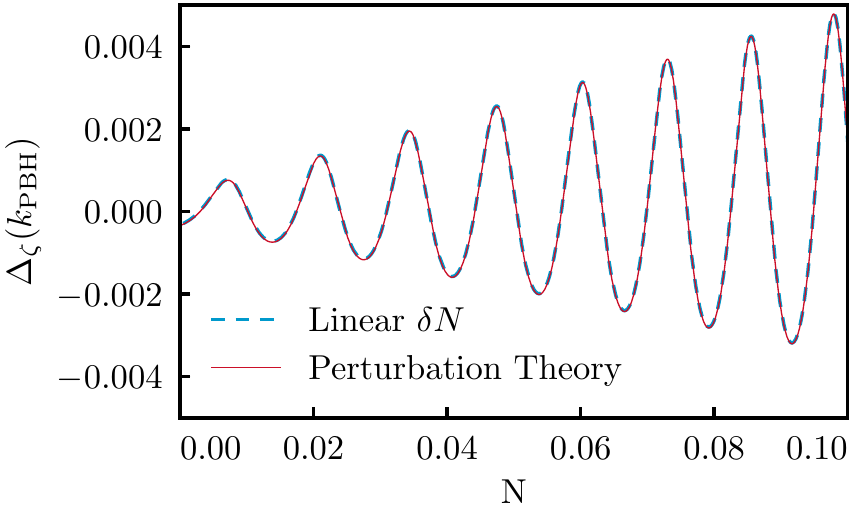}
\caption{The linear perturbation theory calculation including induced radiation fluctuations agrees well with the linearized $\delta N$ result based on Eq.~\eqref{eq:delta_N_linear}, validating our result that energy conservation suppresses curvature fluctuations (see \S\ref{ssec:linear}). 
}
\label{fig:zeta_validation}
\end{figure}

In Fig.~\ref{fig:zeta_validation}, we validate our calculation of $\Delta_\zeta(k_{\rm PBH})$ by also computing $\zeta$ from the linearized $\delta N$ equation \eqref{eq:delta_N_linear}. The $\delta N$ result relies solely on the behavior of the background equations and thus is an independent check on the rather involved perturbation theory calculations. The $\delta N$ result agrees closely with our perturbation theory calculation and confirms that the induced radiation perturbation is crucial in this mechanism. This test would fail if the radiation compensation in
Eq.~(\ref{eq:radcomp}) were omitted as in Refs.~\cite{Espinosa:2017sgp,Espinosa:2018eve,Espinosa:2018euj,Gross:2018ivp}.

$\zeta$ is not conserved after reheating, and in particular it oscillates due to the changing nonadiabatic pressure induced as the Higgs oscillates.  Even though oscillations in $\delta\rho_h$ are relatively small, the
initial near cancellation between $\delta\rho_h$ and $\delta\rho_{\rm r}$ make them prominent in $\zeta$. Moreover, deviations of the Higgs potential from a simple quadratic with a 
temperature-dependent mass make these oscillations grow in time.
We discuss these effects in detail in App.~\ref{app:nonadiabatic}.
So long as the Higgs decay time, generally of order an $e$-fold, is larger than the oscillation timescale, it is the cycle-averaged $\zeta$ that matters. Because of the initial outgoing trajectory of the Higgs, the instantaneous value of $|\zeta|$ at $N=0^+$ is always larger than the cycle average of the first oscillations. 

To the extent that the cycle-averaged Higgs energy redshifts as radiation, the cycle-averaged value of
$\zeta$ will be conserved. However, the nonquadratic terms in the Higgs potential also cause deviation from this behavior which leads the near cancellation between the Higgs and radiation energy densities gradually to break down.

In particular, Higgs perturbations redshift slightly slower than radiation on the cycle average,\footnote{The rate  at which Higgs perturbations redshift \eqref{eq:redshifting_perts} is different from the rate at which the background Higgs redshifts,
\begin{equation}
\langle \rho_h \rangle \propto a^{-4 -3 \Delta \widebar{w}},
\end{equation}
with $\Delta \widebar{w} \sim -0.002$. This means that there is also internal nonadiabatic stress in the
Higgs field itself and so the cycle-averaged $\zeta_h^k$ would also evolve.}
\begin{equation}
\label{eq:redshifting_perts}
\langle \delta \rho_h^k\rangle \propto a^{-4 - 3 \Delta w},
\end{equation}
with $\Delta w \equiv \langle w \rangle - 1/3 \sim -0.004$.
Therefore the cancellation between the radiation piece and the Higgs piece gradually becomes undone, 
\begin{equation}
\zeta^k = - \frac{\delta \rho^k_{\rm r} + \delta \rho^k_h}{\rho_{\rm tot}'} \sim - \frac{\delta \rho^k_{\rm r} + \delta \rho^k_h}{\rho_{\rm r}'} = \zeta^k_{\rm r} + \frac{\delta \rho^k_h}{4 \rho_{\rm r}},
\end{equation}
and $\langle \zeta^k \rangle$ grows gradually. 

Once the Higgs piece dominates, the cycle average becomes
\begin{equation}
\label{zetagrow}
\langle \zeta^k \rangle (N) \sim - \frac{3}{4} \frac{\delta \rho_h^k(0^+)}{\rho_{\rm r}(0^+)} \Delta w \times N \sim 10^{-3} N.
\end{equation}
Thus the curvature grows to $\O{(10^{-1})}$ only on a timescale
\begin{equation}
\label{eq:many_efolds}
\Delta N \sim 100 \text{ $e$-folds} .
\end{equation}
The Higgs must decay to radiation well before this, and therefore the curvature perturbations cannot become large enough in this scenario to form PBHs.

Note that the details of the postreheating evolution and in particular the decancellation rate derived here do depend on the radiation-Higgs split in \eqref{eq:energyconback} through the temperature dependence of the thermal potential \eqref{eq:VT}. However, conservation of total energy on the timescale of Higgs oscillations imposes that the total curvature can grow only on the $e$-fold timescale, and only due to deviations in the redshifting rate of the different components. Therefore the qualitative result that this growth will take many $e$-folds is robust to our specific implementation here.

In summary, under the assumption that linear theory is valid
through reheating, the Higgs instability mechanism falls far short of being able to form PBHs as the
dark matter.  Models that were previously thought to achieve the required $R=0.1$ in fact produce
$R \lesssim 10^{-3}$ once the radiation density perturbations required by energy conservation at 
reheating are properly accounted for. 

\subsection{Nonlinear Conversion}
\label{ssec:nonlin}

In \S\ref{ssec:linear}, we computed curvature fluctuations assuming that the Higgs perturbations remain linear 
through reheating.  In fact, the Higgs instability induces a breakdown of linear theory when the background position of the Standard Model Higgs at the end of inflation $\hend$ is close to the maximum rescue scale $\hrescue$.  

This breakdown can be seen immediately from Fig.~\ref{fig:rho}. With $\delta N = \zeta_h \sim \pm1$, a typical outwardly perturbed region of the Standard Model Higgs field crosses $\hrescue$ during inflation, gains exceedingly large negative energy and will inevitably backreact on the background trajectory. Reheating will be disrupted, the perturbed Higgs will not be rescued from the unbounded vacuum, and our universe will be destroyed.  Even for smaller $\delta N$ which do not cross $\hrescue$,  the perturbed Higgs energy density is not well represented by the linear Taylor expansion (\ref{eq:linear_jump})  around the background value
due to the extremely rapid evolution of $\rho_h$.

In terms of field interactions, linear theory itself also reveals its own breakdown. At the end of inflation, an order unity $\zeta_h$ leads to a rms Higgs fluctuation of roughly
\begin{equation}
\delta h_{\rm end} \simeq h'_{\rm end} \simeq -\frac{1}{3 H^2} \lambda \hend^3,
\end{equation}
where we have used the Higgs slow-roll approximation throughout. With $\hend\sim10^3 H$ and $|\lambda|\sim 10^{-2}$ we have 
\begin{equation}
\delta h_{\rm end} \simeq 10^7 H,
\end{equation}
which as we have seen is orders of magnitude larger than the distance between $\hend\simeq 1200H$ and $\hrescue \simeq 2400H$. Moreover, the potential interaction ratio \eqref{eq:potential_linearity}  is   
\begin{equation}
\frac{1}{2} \delta h \frac{V_{,hhh}}{V_{,hh}} \simeq  \frac{\delta h}{4 h} \simeq  10^4 \gg 1,
\end{equation} 
and thus field fluctuations interact. 
Nevertheless, this breakdown has no effect on our previous computation of $\Delta_{\zeta_h}^2$ during inflation since $\zeta_h$ is conserved nonlinearly so long as $H \sim$ const.

Though it was not phrased in terms of a breakdown of linear theory, Ref.~\cite{Gross:2018ivp} noted that the Standard Model Higgs is generally not rescued in this scenario. It was argued in Ref.~\cite{Espinosa:2018euj} that the background $\hend$ can be placed near $\hrescue$ while multiverse and anthropic considerations justify tuning the local Higgs field at the end of inflation such that ${\rm Max}\left[h(\vec{x}) \right] < \hrescue$ everywhere. However, tuning $\delta h (\vec{x}) $ at the end of inflation is  equivalent to tuning $\delta N = \zeta_h$ to be small at the end of inflation and so it directly tunes away the ability to form PBHs as we shall now show.

\begin{figure}[t]
\includegraphics{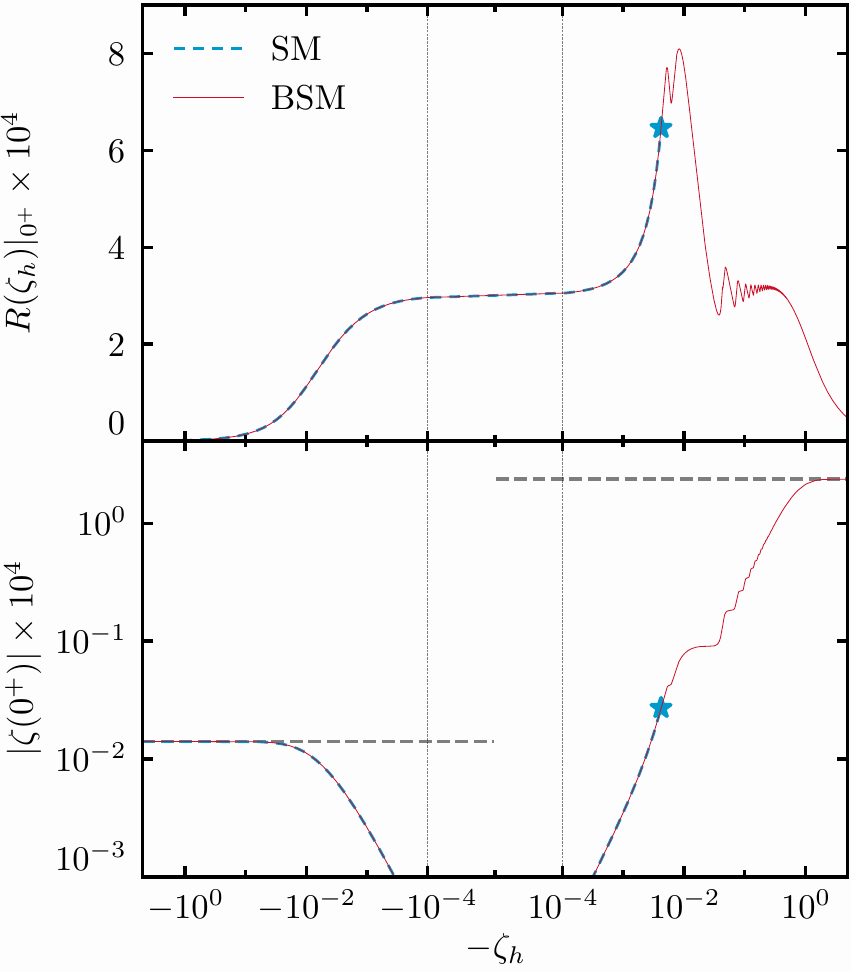}
\caption{The  nonlinear mapping of the inflationary $\zeta_h$ to the postreheating $\zeta (0^+) = \left.R(\zeta_h)\right\vert_{0^+} \zeta_h$. Fluctuations outward,
toward the instability, correspond to $-\zeta_h>0$. The horizontal axis scale is linear between $\pm 10^{-4}$ and logarithmic elsewhere. 
$R$ deviates from the linear theory 
value $\simeq 3\times10^{-4}$ for fluctuations larger
than about $\pm 10^{-3}$. The horizontal dashed lines indicate analytic saturation values
\eqref{eq:BSMsatin} and \eqref{eq:BSMsatout}.
For the SM Higgs, the maximal $\zeta$ satisfying the rescue condition \eqref{eq:rescue} is marked with a star.
For the BSM Higgs,  $\zeta$ saturates to a maximum.   Neither value is large enough to form PBHs
}
\label{fig:zeta_zetah}
\end{figure}

In Fig.~\ref{fig:zeta_zetah} we show $\zeta(0^+)$ as a function of $\zeta_h$ as computed using the nonlinear $\delta N$ formalism 
using Eq.~(\ref{eq:nonlin_jump2}) and Fig.~\ref{fig:rho}.
Linear theory holds for small enough inflationary $\left\vert\zeta_h\right\vert \lesssim 10^{-3}$, but breaks down for perturbations of the typical amplitude produced during inflation. 

Large inward perturbations away from the instability, shown on the left-hand side of Fig.~\ref{fig:zeta_zetah}, 
saturate to a constant $\zeta(0^+)$ that is independent of $\zeta_h$. These uphill kicks produce
a local Higgs energy density at $N=0^-$ that has a much smaller magnitude than its background value as can be
seen in Fig.~\ref{fig:rho}.
Using Eq.~\eqref{eq:nonlin_jump}, the left-hand side saturation can therefore be written as
\begin{equation}
\zeta^{\rm in}_{\SM} (0^+) =  \frac{- \bar{\rho}_h(0^-)}{12 H^2} \simeq +1.4 \times 10^{-6},
\end{equation}
which we show as a horizontal dashed line on the left-hand side in Fig.~\ref{fig:zeta_zetah}.

Outward perturbations of the SM Higgs toward the instability, shown on the right-hand side of Fig.~\ref{fig:zeta_zetah}, are enhanced relative to linear theory. This is because the amplitude of the energy density of the Higgs shown in Fig.~\ref{fig:rho} grows much faster than expected from a linear approximation. The largest outward perturbations that satisfy the rescue condition \eqref{eq:rescue} produce a curvature
\begin{equation}
\zeta_{\SM}^{\rm out} (0^+) = \frac{\rho_h (\hrescue) - \bar{\rho}_h(0^-)}{12 H^2} \simeq -2.7 \times 10^{-6}.
\end{equation}

Despite the enhancement of the Higgs perturbation relative to linear theory, $\zeta_{\SM}^{\rm out}$ evolves after inflation much like the linear theory $\zeta$ computed in \S\ref{ssec:linear}. The cycle average of $\zeta_{\SM}^{\rm out}(N)$ is smaller than $\zeta_{\SM}^{\rm out} (0^+)$.
So long as the Higgs redshifts like radiation after inflation, this value is then conserved.   Nonlinear evolution
does not change the conclusion of linear theory on PBHs with the SM potential. 

So far in this nonlinear calculation we have kept the background trajectory of the Higgs fixed. We might wonder whether a different background Higgs trajectory, at fixed $H$, can achieve a larger value for the saturating $\zeta$. Moving the background away from the instability suppresses $\bar{\rho}_h(0^-)$, and so the largest curvature perturbation which can be produced in the SM, maximized over the choice of background trajectory, is 
\begin{equation}
\zeta_{\SM}^{\rm max} (0^+) = \zeta_{\SM}^{\rm out}(0^+) - \zeta_{\SM}^{\rm in}(0^+) \simeq -4.1 \times 10^{-6},
\end{equation}
which is still insufficient to form PBHs.  
Therefore since
$\zeta$ is uniquely determined by $\zeta_h$ in this way,
anthropically tuning away $\zeta_h({\vec x})$ or setting it to the edge of rescuability forbids
the Standard Model Higgs from forming  PBHs in sufficient abundance to be the dark matter.
  
Ref.~\cite{Espinosa:2018euj} proposed that the mechanism could function with a BSM potential derived from the addition of a massive scalar as detailed in \S\ref{sec:mechanism}. The massive scalar adds a wall in the potential between the field value where the background Higgs ends inflation $\hend$ and the maximum rescuable point $\hrescue$, preventing the local Higgs from reaching parts of the potential from which it cannot be rescued by thermal uplift at reheating. 

When CMB and PBH modes cross the horizon, the BSM potential behaves like the SM potential and as we have seen in \S\ref{sec:inflation} this means that it generates larger inflationary $\zeta_h$ perturbations on CMB scales than on PBH scales. In addition, so long as the background trajectory never encounters the wall, this model behaves like the SM in linear theory and yields $R\ll 0.1$ as shown in \S\ref{ssec:linear}.

However, whereas typical regions with outward field fluctuations were not rescued in the SM, in the BSM case such regions oscillate in a new minimum of the potential during inflation and then can be safely rescued at reheating. 

To evolve fluctuations through this highly nonlinear process, we again use the nonlinear $\delta N$ formalism described in \S\ref{ssec:deltaN}, just as in the Standard Model case, and the BSM results for $\zeta(0^+)$ are also shown in Fig.~\ref{fig:zeta_zetah}. We compute results using the representative parameter set for the BSM scalar described in \S\ref{sec:mechanism}, and we will later show how our results scale with different choices of model parameters. 

Inward perturbations of the BSM Higgs act just like inward perturbations in the SM and thus again lead to the same saturation
\begin{equation}
\label{eq:BSMsatin}
\zeta^{\rm in}_{\BSM}(0^+) = \zeta^{\rm in}_{\SM}(0^+) \simeq +1.4 \times 10^{-6}.
\end{equation}

Large outward Higgs perturbations hit the BSM potential wall, become trapped in the new minimum at $h\sim m_s$,
lose their kinetic energy, and end inflation with a potential dominated Higgs with energy $V_{\rm min}$.
This leads to a saturating curvature
\begin{equation}
\label{eq:BSMsatout}
\zeta^{\rm out}_{\BSM}(0^+) = \frac{V_{\rm min} - \bar{\rho}_h(0^-)}{12 H_{\rm end}^2} \simeq -2.4 \times 10^{-4},
\end{equation}
which we show with a horizontal dashed line on the right-hand side in Fig.~\ref{fig:zeta_zetah}. 
This value is still too small to form enough PBHs to be the dark matter. 

For perturbations which do not fully saturate this limit, $\zeta(0^+)$ has a stepped behavior and $R(0^+)$ an oscillatory one as depicted in Fig.~\ref{fig:zeta_zetah}. These features correspond to the energy density oscillations for the BSM Higgs seen in Fig.~\ref{fig:rho}, induced because the Higgs has large oscillations around the potential minimum before settling on the $e$-fold timescale. Note that the approximate equality of the linear theory $R(0^+)$ and the nonlinear $R(0^+)$ corresponding to $\vert{\zeta_h}\vert \simeq 0.1$ is a coincidence: changes to the background position change the linear theory $R(0^+)$ while leaving $R(0^+)$ on this brief plateau fixed.

Once more we might wonder if a different choice of background trajectory could enhance the saturation value, but in fact the maximum over background trajectories can again be computed as
\begin{equation}
\zeta_{\BSM}^{\rm max}(0^+) =  \zeta_{\BSM}^{\rm out}(0^+) - \zeta_{\BSM}^{\rm in}(0^+) \simeq \zeta^{\rm out}_{\BSM}(0^+),
\end{equation}
and therefore PBHs cannot be formed for our fiducial BSM potential no matter the position of the background or the size of initial fluctuation.

Note that so far we have only computed $\zeta(0^+)$ for BSM. We should check whether $\zeta$ evolves significantly after $N=0^+$. We do so again with the $\delta N$ formalism, using the full Eq.~\eqref{eq:delta_N}. Since inward fluctuations lead to a negligible $\zeta(0^+)$, we can select a typical outward field fluctuation with $\zeta_h(\vec{x}) = \Delta_{\zeta_h}(k_{\rm PBH}) \simeq -1$ as an example.

\begin{figure}[t]
\includegraphics{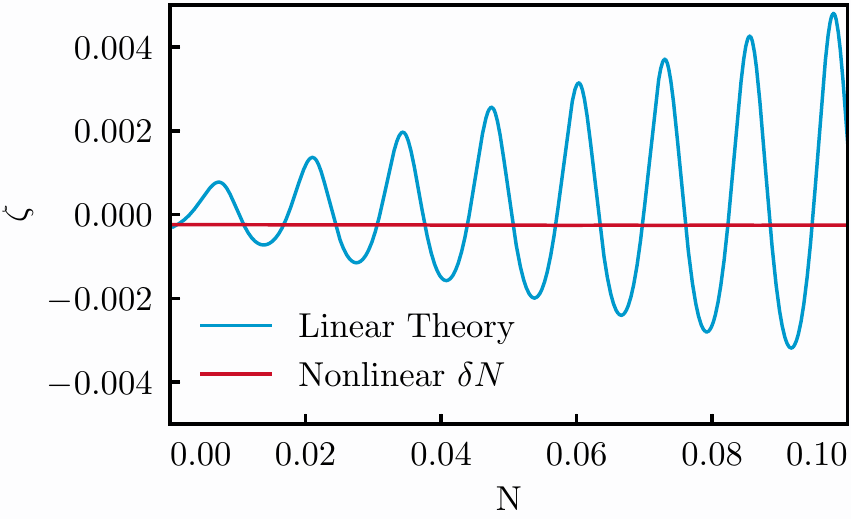}
\caption{Curvature evolution after reheating. The  BSM nonlinear $\delta N$ result for the local $\zeta(\vec{x})$, from an example inflationary 
$\zeta_h(\vec{x}) = \Delta_{\zeta_h}(k_{\rm PBH})$, is compared to the linear theory approximation $\Delta_{\zeta}(k_{\rm PBH})$ of Fig.~\ref{fig:zeta_validation}. The nonlinear evolution of $\zeta$ after reheating is too small to form PBHs (see \S\ref{ssec:nonlin}). 
}
\label{fig:zeta_nonlin}
\end{figure}

We show this case in Fig.~\ref{fig:zeta_nonlin}. As expected from Fig.~\ref{fig:zeta_zetah}, the nonlinearly evolved $\zeta$ is small, comparable in amplitude  to the linear $\zeta(0^+)$ but not in its evolution.   In fact $\zeta$ evolves negligibly after $N=0^+$ 
and we can robustly conclude that  PBHs are generically not formed nonlinearly in this case. 

This lack of nonlinear evolution can be explained by the difference in the impact of
uplift on the perturbations.
In linear theory the small amplitude of $\zeta$ resulted from large cancellations between the Higgs and radiation perturbations due to energy conservation and
the large impact of uplift. Nonlinearly the impact of uplift is much smaller, bounded by the BSM modification, and so the Higgs energy density fluctuations after reheating are no longer
as dominated by the uplift contribution.   In particular,
\begin{equation}
\frac{\rho_h(0^+,\vec{x}) - \bar{\rho}_h(0^+)}{\rho_h(0^-,\vec{x}) - \bar{\rho}_h(0^-)} \ll \left.\frac{\delta \rho_h (0^+)}{\delta \rho_h(0^-)}\right\vert_{\rm linear},
\end{equation}
where the right-hand side is in linear theory. Therefore the cancellation with radiation is less dramatic than in linear theory. As discussed in \S\ref{ssec:linear}, the cancellation and subsequent decancellation are responsible for the linear theory oscillations and slow secular drift~\eqref{zetagrow}, and therefore all these effects are suppressed in the nonlinear case. For the same reason, the details of the split of the total energy density into Higgs and radiation pieces are also less important for the nonlinear curvature. Our
nonlinear results therefore essentially depend only on energy conservation during reheating.

Finally, the fiducial BSM potential used to compute the results of Fig.~\ref{fig:zeta_zetah} was constructed according to the specifications of Ref.~\cite{Espinosa:2018euj}: it uplifts the Standard Model potential somewhere between $\hend$ and $\hrescue$. We might wonder whether PBHs could be formed by optimizing the position of the uplift so that it as close as possible to $\hrescue$, maximizing the criticality of the scenario. 

The maximum position of $m_s$ will be just before $\hrescue$. This leads to a maximum curvature for this entire scenario of 
\begin{equation}
\label{eq:deltaN_max_BSM}
\textrm{Max}[\zeta_{\BSM}^{\rm max}] =  \frac{V(\hrescue)}{12 H_{\rm end}^2}.
\end{equation}
Using the approximate maximum rescuable field value \eqref{eq:hrescue} and $\lambda = \lambda^{\SM} \sim - 0.007$, we have
\begin{equation}
V(\hrescue^{(1)}) = \frac{1}{4} \lambda^{\SM} \left(\hrescue^{(1)}\right)^4 = -0.012 H_{\rm end}^2,
\end{equation}
which yields
\begin{equation}
\label{eq:zeta_max_max}
\zeta^{\rm max}_{\BSM} \simeq -1.0 \times 10^{-3},
\end{equation}
which depends on $H$ only through the logarithmic evolution of $\lambda^{\SM}$ evaluated at $\hrescue$. This estimate is in good agreement with the computation using the exact value of $\hrescue$ which yields $\zeta^{\rm max}_{\BSM} = -8.2 \times 10^{-4}$.

Therefore no matter the size of inflationary Higgs perturbations, the position of the background Higgs, the SM or BSM nature of the Higgs potential, the position of the BSM wall, or the Hubble rate, this mechanism does not produce perturbations large enough to form PBHs in sufficient abundance to be the dark matter.

Moreover, the largest possible curvature perturbations produced by this model, Eq.~\eqref{eq:zeta_max_max}, are so small that the second-order gravitational waves predicted by Ref.~\cite{Espinosa:2018eve} will be undetectable with LISA.

\begin{figure}[t]
\includegraphics{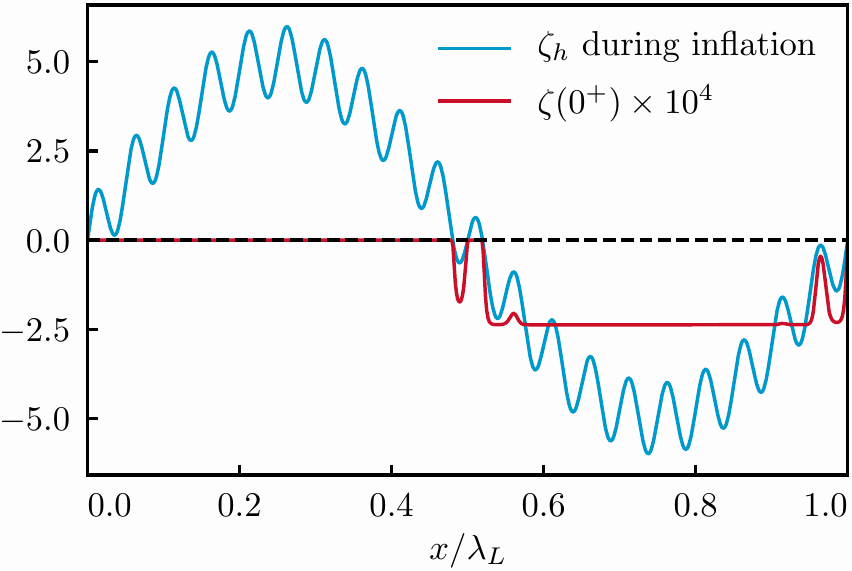}
\caption{The real-space postinflationary curvature field $\zeta$ (red) produced by the nonlinear transformation of the inflationary $\zeta_h$ (blue). Perturbations on small scales are suppressed when they occur where the long-wavelength $\lambda_L$ mode has saturated the curvature field, and therefore CMB scale perturbations are larger than PBH scale perturbations fully nonlinearly in this mechanism (see \S\ref{ssec:nonlin}). 
}
\label{fig:zeta_real_space}
\end{figure}

We can also ask what happens to CMB scale fluctuations nonlinearly in this model. We showed in \S\ref{sec:inflation} that during inflation $\Delta^2_{\zeta_h}(k_{\rm CMB})  > \Delta^2_{\zeta_h}(k_{\rm PBH})$. In Fig.~\ref{fig:zeta_real_space}, we show the effect of the highly nonlinear local transformation of $\zeta_h$ to $\zeta$ shown for the BSM potential in Fig.~\ref{fig:zeta_zetah} on a cartoon realization of the curvature field on a constant Higgs density surface during inflation.

For visualization purposes, we have generated two modes apart by a factor of only $20$ in scale rather than the $\sim35$ $e$-folds which separate the CMB and PBH modes. We see that fluctuations on the long-wavelength scales cause a saturation of the short-wavelength fluctuations, and therefore fully nonlinearly we have that perturbations are larger on long wavelengths than on small wavelengths in this model as long as this is true of the inflationary $\zeta_h$ itself. Therefore if the reheating scenario is changed somehow to achieve large PBH scale fluctuations, the CMB scale fluctuations will still be larger than PBH modes unless the functional form of the transformation shown in Fig.~\ref{fig:zeta_zetah} is radically altered.

Beyond the specific motivation of PBH dark matter formation, we can now return to the question of whether  Higgs instability is compatible with the small curvature fluctuations observed in the CMB.

For the SM Higgs, suppressing $\zeta^{\rm out}_{\SM}$ requires placing the background $\hend$ far from $\hrescue$. Specifically, since the typical CMB scale perturbation has $|\zeta_h| \sim 5$, $\hend$ should be moved at least $\sim5$ $e$-folds backward along its trajectory.

For the BSM potential, with the scalar mass set near $\hrescue$ to ensure that no regions ever fall into the unrescuable region, there are two situations which are compatible with the CMB and one which is not.

If $\hend$ does not approach $\hrescue$, then we return to the linear theory, Standard Model result, where Higgs fluctuations on CMB scales do not lead to large curvature fluctuations since
$\rho_h/H^2$ decreases sharply in Fig.~\ref{fig:rho} and predicts $\zeta$ through Eq.~(\ref{eq:nonlin_jump2}).

Conversely, if the Higgs travels far down the unstable region early in inflation, 
the Higgs becomes uniform in the potential well during inflation and thus leads to no curvature perturbations after inflation. In this region as well the Higgs instability is compatible with the CMB.

It is only in the region of parameter space near our fiducial model, where the background Higgs $\hend$ approaches but does not reach the minimum induced by the BSM massive scalar near $\hrescue$, that Higgs fluctuations on CMB scales can be converted to curvature fluctuations which are large enough to disturb the CMB. 

To avoid this possibility completely, one should set the mass $m_s$ of the scalar field to be slightly smaller than $\hrescue$. In particular to achieve $|\zeta|\lesssim10^{-5}$, using Eqs.~\eqref{eq:deltaN_max_BSM} and \eqref{eq:zeta_max_max}, one requires
\begin{equation}
\frac{m_s}{\hrescue} \lesssim \left(\frac{1}{100}\right)^{\frac{1}{4}} \lesssim \frac{1}{3}.
\end{equation}

In summary, it is only a special class of Higgs criticality scenarios where the parameters are arranged so
that the regions of the universe fluctuate near the edge of rescuable instability which would
be testable in the CMB and even that class cannot form PBHs as the majority of the dark matter,
nor generate second-order gravitational waves at an amplitude detectable with LISA.

\section{Conclusions}
\label{sec:conclusions}

We have definitively shown that the dark matter is not composed exclusively of primordial black holes produced by the collapse of density perturbations generated by a spectator Higgs field during inflation.

While a spectator Higgs field evolving on the unstable side of its potential can generate large Higgs fluctuations on PBH scales, even larger Higgs fluctuations are produced on CMB scales. 
This result is obtained using linear perturbation theory during inflation, which we show holds even though the CMB modes cross the horizon at an epoch when the Higgs' per $e$-fold classical roll is smaller than its per $e$-fold stochastic motion, because the stochastic motion is incoherent and does not backreact on the Higgs background. 

Inflation ends well after all relevant modes have crossed the horizon, and when reheating occurs the Higgs potential is uplifted by the interaction between the Higgs and the thermal bath. If the Higgs is rescued from the unstable region by this thermal uplift then the Higgs redshifts as, and eventually decays to, radiation. The CMB and PBH modes are superhorizon at these epochs and evolve in the same way through these processes. Therefore if the Higgs fluctuations are converted into sufficiently large curvature perturbations such that enough PBHs were produced to explain the dark matter, CMB constraints would necessarily be violated.

In fact though, a sufficient abundance of PBHs is never produced and CMB constraints are only violated in cases of near criticality. We first showed that this is true under the assumption that linear theory holds through reheating, where we correct an error in local energy conservation made in the literature.

We then showed that linear theory is violated because the model requires the Higgs to be as close as possible to the maximum value beyond which it cannot be rescued at reheating. This criticality condition leads typical perturbations to evolve nonlinearly, and using the nonlinear $\delta N$ formalism we also show that the Standard Model Higgs, regardless of fine-tuning or anthropic arguments, can never produce enough PBHs to be the dark matter. Modifying the Higgs potential at large field values can eliminate fine-tuning or anthropic issues, but cannot enhance curvature perturbations significantly enough to explain the dark matter. 

\acknowledgments

	We thank Andrew J. Long for fruitful discussions along with Peter Adshead, Jose Mar\'{i}a Ezquiaga, Rocky Kolb, and Lian-Tao Wang for helpful comments.
	SP and WH were supported by U.S.\ Dept.\ of Energy contract DE-FG02-13ER41958 and the Simons Foundation.  SP was additionally supported by the Kavli Institute for Cosmological Physics at the University of Chicago through grant NSF PHY-1125897 and an endowment from the Kavli Foundation and its founder Fred Kavli.
	HM was supported by Japan Society for the Promotion of Science (JSPS) Grants-in-Aid for Scientific Research (KAKENHI) No.\ JP17H06359 and No.\ JP18K13565.

\appendix

\section{Formation of inflationary background by superhorizon stochastic modes}
\label{app:nonini}

In \S\ref{ssec:background} we showed that the Higgs field fluctuations are linearizable around a slow-roll background when all observationally relevant scales crossed the inflationary horizon. In \S\ref{ssec:attini}, we used that background to compute Higgs fluctuations during inflation. Here, we will show that such a background can be produced in the stochastic inflation formalism, and that such a scenario yields results consistent to those of \S\ref{ssec:attini}.

In the stochastic picture, our background on the unstable side of the potential is formed by Higgs fluctuations which crossed the horizon before $N\sim-60.$ At horizon crossing, 
each such fluctuation has an amplitude
\begin{equation}
\label{eq:stochastic_kick}
\delta h \simeq \frac{H}{2 \pi},
\end{equation}
and a velocity
\begin{equation}
\label{eq:stochastic_vel}
{\delta h}' \simeq \frac{H}{2 \pi}.
\end{equation}

Our background was formed when a cumulative series of such stochastic kicks pushed our Hubble patch to the unstable side of the barrier. We can reabsorb these early stochastic kicks into a background field that is spatially homogeneous on our Hubble patch, while subsequent stochastic kicks lead to spatial perturbations on smaller scales which we addressed in \S\ref{sec:inflation}. 

When absorbed into a FLRW background in our Hubble volume, a large-scale stochastic kick to the Higgs imparts a velocity 
\begin{equation}
\Delta v =  \frac{H}{2\pi}.
\end{equation}
where $v \equiv d h / d N$. The background then obeys the equation of motion \eqref{eq:higgs_bg_eom} and when the potential derivative term is negligible the velocity decays according to 
\begin{equation}
\label{eq:vel_decay_FLRW}
\frac{d \ln v}{d N} = -3 .
\end{equation}
However, the velocity of a superhorizon mode in the absence of classical roll
can be deduced from the de Sitter mode function \eqref{eq:bunch-davies}, which yields
\begin{equation}
\label{eq:vel_decay_true}
\frac{d\ln {\delta h}'}{d N} = -2 .
\end{equation}
The difference between these decay rates is a breaking of the separate universe condition. If we reabsorb the superhorizon stochastic kicks into a new FLRW background we make a small error.

Nonetheless, the superhorizon velocity decay is exponential and the field velocity is rapidly dominated by classical roll.
Only kicks which occurred just prior to $N\sim-60$ contribute to the velocity of our background. The cumulative stochastic kicks therefore impart to our background an initial velocity 
\begin{equation}
\label{eq:horizon_velocity}
h'_{60} \sim \frac{H}{2 \pi},
\end{equation}
where $\sim$ indicates a multiplicative factor of order unity.
As described in \S\ref{ssec:background}, kicks after $N=-60$ do not backreact on the background and we treat those as linear perturbations. Therefore after $h_{60}$ the initial velocity decays until it becomes less important than the potential derivative term and our background has reached the attractor solution.

The numerical solution of the FLRW background Eq.~\eqref{eq:higgs_bg_eom} with the initial condition Eq.~\eqref{eq:horizon_velocity} shows that the initial position of the field should be shifted by
\begin{equation}
\Delta h_{60} \sim 0.04 H
\end{equation}
away from the instability in order to achieve the same $\hend$ as in the attractor initial velocity case. 
Therefore, the nonattractor initial velocity~\eqref{eq:horizon_velocity} has little impact on the position of our background. This is because the nonattractor phase lasts just a small fraction of an $e$-fold, which can be seen as follows. At $h_{60}$, the Higgs potential slope is
\begin{equation}
- \frac{1}{3 H^2} \left.V_{,h}\right\vert_{h_{60}} \sim \frac{1}{3} \times \frac{H}{2 \pi},
\end{equation}
which means that from Eq.~\eqref{eq:vel_decay_FLRW} the initial velocity for the FLRW background becomes less than the potential slope in $(\ln 3)/3 \simeq 0.4$ $e$-folds. If we had used the correct superhorizon velocity decay Eq.~\eqref{eq:vel_decay_true}, then we would have found that the initial velocity decays in $(\ln 3)/2 \simeq 0.5$ $e$-folds. Therefore this error is not important for small-scale modes that crossed the horizon after this epoch if we assume that the background in our Hubble volume is
established by stochastic kicks.

\section{Superhorizon Curvature Evolution}
\label{app:nonadiabatic}

In this appendix, we discuss how and why curvature perturbations can evolve on superhorizon scales both during and after inflation in this scenario.

Local conservation of the stress-energy of a noninteracting fluid $I$
\begin{equation}
\nabla^{\mu} T^I_{\mu \nu} = 0,
\end{equation}
yields at the perturbation level a continuity equation and a Navier-Stokes equation. For our purposes, $I$ here will either denote the Higgs fluid or the total fluid, during or after inflation.

The continuity equation on a surface of uniform energy density of $I$ leads to a conservation equation for the curvature perturbation $\zeta_I$ on that surface. On superhorizon scales, the conservation equation takes the simple form
\begin{equation}
\label{eq:conservation}
{\zeta_I}' = - \frac{\delta p_I^{\rm NA}}{\rho_I+p_I},
\end{equation} 
where $p_I$ is the fluid pressure, $\rho_I$ is the energy density, 
and $\delta p_I^{\rm NA}$ is the nonadiabatic pressure of the fluid (see, e.g., Ref.~\cite{Hu:2004xd} for notation). The nonadiabatic pressure is the pressure perturbation on a surface of uniform density of $I$.

The nonadiabatic pressure on the right-hand side of the superhorizon conservation equation~\eqref{eq:conservation} can be computed either by directly studying local variations in pressure on a uniform density surface, or by subtracting the adiabatic pressure through
\begin{equation}
\delta p_I^{\rm NA} = \delta p_I - \frac{\dot p_I}{\dot\rho_I} \delta \rho_I 
\label{eq:pNA}
\end{equation}
where $\delta p_I$ and  $\delta \rho_I$ are the pressure and density perturbations of the fluid $I$ in any gauge.  The appearance of nonadiabatic pressure is associated with
having multiple degrees of freedom or clocks so that the local density no longer uniquely
defines the local pressure. For example for a scalar field,  if the kinetic
energy were not uniquely specified by the potential energy, i.e.\ the field position, then
there is nonadiabatic stress. Likewise if $I$ is a composite of two systems with
differing equations of state then $\rho_I$ does not uniquely define $p_I$.  We shall see that both mechanisms are operative for the Higgs instability calculation.

Meanwhile the left-hand side of the superhorizon conservation equation~\eqref{eq:conservation}, $\zeta_I'$ can be computed by taking a derivative of the curvature perturbation computed in the $\delta N$ approach (see \S\ref{ssec:deltaN} for an overview of $\delta N$), or by taking a derivative of the curvature perturbation obtained by transformation from the spatially flat gauge. For the Higgs, the Fourier mode of the flat gauge field perturbation satisfies the Klein-Gordon equation
\begin{align*}
\label{eq:linKGwithmetric}
 &\left(\frac{d^2}{d \eta^2} + 2 \frac{\dot a}{a} \frac{d}{d \eta} +  k^2   \right) \delta h^k + a^2  \delta V^k_{,h} \\
 &= (\dot {A}- k B) \dot{h} - 2 a^2 A V_{,h}. \numberthis
\end{align*}
The terms appearing on right-hand side of the Klein-Gordon equation~\eqref{eq:linKGwithmetric} involve
the $k$-modes $A$ of the lapse perturbation and $B$ of the scalar shift perturbation, in the spatially flat gauge, which we neglected in \eqref{eq:linKG}. These terms take different forms during and after inflation and we shall provide them shortly. We drop their $k$ superscripts for compactness. 

These different approaches allow us to better understand why the curvature evolves on superhorizon scales during inflation, at reheating, and after inflation, and they enable us to check our calculations for self-consistency, including that of various metric and field terms which will appear.

\subsection{During Inflation}
 
During inflation, the Higgs has energy density 
\begin{equation}
\label{eq:background_rho_during}
\rho_h = \frac{1}{2} \frac{\dot{h}^2}{a^2} + V,
\end{equation}
and pressure
\begin{equation}
\label{eq:background_p_during}
p_h  = \frac{1}{2} \frac{\dot{h}^2}{a^2}- V.
\end{equation}
In the flat gauge, the Fourier mode of the Higgs density perturbation is
\begin{equation}
\delta \rho_h^k = \frac{1}{a^2}\left(\dot{h} \dot{\delta h^k} - \dot{h}^2 A\right) + V_{,h} \delta h^k,
\end{equation}
and the pressure perturbation is
\begin{equation}
\delta p_{h}^k = \frac{1}{a^2}(\dot{h} \dot{\delta h^k} - \dot{h}^2 A) - V_{,h} \delta h^k.
\end{equation}
The metric terms in these equations and in the Klein-Gordon equation~\eqref{eq:linKGwithmetric} satisfy the momentum constraint and Hamiltonian (Poisson) constraint equations 
\begin{align*}
\frac{\dot{a}}{a} A &= \frac{1}{2} \dot h \delta h^k + \frac{1}{2} \dot \phi \delta \phi^k, \\
\frac{\dot{a}}{a} k B &= \frac{1}{2} a^2 (\delta \rho_h^k + \delta \rho_\phi^k) +  3 \left(\frac{\dot a}{a}\right)^2 A.\numberthis
\end{align*}
The Klein-Gordon equation \eqref{eq:linKGwithmetric} then becomes 
\begin{align*}
\label{eq:KGmetric}
 &\left(\frac{d^2}{d \eta^2} + 2 \frac{\dot a}{a} \frac{d}{d \eta} +  k^2   \right) \delta h^k + a^2  \delta V^k_{,h} \\ 
 &= \delta h^k \frac{1}{a^2} \frac{d}{d \eta} \left( \frac{a^3}{\dot{a}} \dot{h}^2 \right) + \delta \phi^k \frac{1}{a^2} \frac{d}{d \eta} \left( \frac{a^3}{\dot{a}} \dot{h} \dot{\phi} \right).
\numberthis
\end{align*}
The metric terms contain contributions from the Higgs perturbations themselves as well as inflaton perturbations induced by the metric perturbations in the analogous Klein-Gordon equation for the inflaton. Intrinsic inflaton fluctuations should not be included here since they are uncorrelated with the Higgs and computed separately. 

We have verified numerically that including these metric terms on the right-hand side of the Klein-Gordon equation~\eqref{eq:KGmetric}, including accounting for the induced inflaton perturbations, has no significant effect on the solution for $\delta h^k$ or $\delta \rho_h^k$, which justifies our treatment in \S\ref{ssec:attini}. This can be analytically seen as follows. 

Given that the Klein-Gordon source to the field velocity is the potential slope, we can estimate
\begin{equation} 
\label{eq:hpest}
h' \sim  -c \frac{V_{,h}}{3 H^2}, 
\end{equation}
where $c=1$ for Higgs slow roll and an order unity factor otherwise.
Assuming that the fields in the background evolve on the Hubble time or slower,
the rhs of Eq.~(\ref{eq:KGmetric}) can then be estimated 
as 
\begin{align*}
 &\left(\frac{d^2}{d \eta^2} + 2 \frac{\dot a}{a} \frac{d}{d \eta} +  k^2   \right) \delta h^k + a^2  \delta V^k_{,h} \\ 
 &\simeq \delta h^k \frac{c^2 a^6}{\dot{a}^2} V_{,h}^2 - \delta \phi^k c \frac{a^3}{\dot{a}} V_{,h} \dot{\phi}. \numberthis
\end{align*}

The $\delta h^k$ metric term on the rhs can be compared to the nominal $\delta V^k_{,h} = \delta h^k V_{,hh}$ term on the left-hand side. Using the analytic form for the Higgs potential, $V = \lambda h^4 / 4$, with $\lambda$ only logarithmically varying, shows that the metric $\delta h^k$ term is suppressed relative to the nominal term as
\begin{equation}
\frac{c^2 V_{,h}^2}{3 H^2 V_{,hh}} 
\simeq \frac{c^2}{9} \frac{\lambda h^4}{H^2} = \frac{4 c^2}{9} \frac{V}{H^2} \ll 1
\end{equation}
where the last inequality follows because the Higgs is a spectator during inflation.

The second metric term, proportional to $\delta \phi^k$, is further suppressed relative to the $\delta h^k$ metric term because $\delta \phi^k$ is sourced by the Higgs metric term itself. 
In particular, solving the inflaton Klein-Gordon equation with the Higgs metric source we find
\begin{equation}
\delta \phi^k  \sim \delta h^k h' \phi',
\end{equation}
and therefore in the Higgs Klein-Gordon equation the inflaton metric term becomes suppressed relative to the Higgs one by a factor
\begin{equation}
\phi'^2 \sim \epsilon_H \lesssim 1.
\end{equation}
Therefore all metric terms in the Klein-Gordon equation~\eqref{eq:KGmetric} for the Higgs can be neglected when the Higgs is a spectator. 
This justifies the approximated Klein-Gordon equation~\eqref{eq:linKG} and the subsequent analysis based on it.  As shown in \S\ref{ssec:attini}, 
the curvature perturbation on uniform Higgs density (UHD) surfaces during inflation then obeys \eqref{eq:sup}, i.e.\
\begin{equation}
\label{eq:sup2}
\frac{{\zeta_h^k}'}{\zeta_h^k} = -2\e,
\end{equation}
on superhorizon scales at leading order in the Hubble slow-roll parameter $\e$. 
This superhorizon evolution is important for the estimation~\eqref{eq:zeta_k_h_comparison} of the relative amplitude of $\Delta^2_{\zeta_h}$ on CMB and PBH scales.

We can alternately derive the superhorizon evolution~\eqref{eq:sup2} by using the nonadiabatic pressure relation~\eqref{eq:pNA} or by using the $\delta N$ formalism~\eqref{eq:delta_N_linear}.
Let us first focus on the nonadiabatic pressure.
On a UHD surface, the Higgs energy density fluctuation vanishes
\begin{equation}
\delta (\rho_h)_{\rm UHD} \equiv 0 = \frac{1}{2} \delta (H^2 h'^2)_{\rm UHD} + V_{,h} \delta(h)_{\rm UHD},
\end{equation}
where we have again assumed metric perturbations are negligible. The nonadiabatic pressure~\eqref{eq:pNA} is then
\begin{align*}
\delta p^{\rm NA}_h \equiv \delta (p_h)_{\rm UHD}  &= \frac{1}{2} \delta (H^2 h'^2)_{\rm UHD}  - V_{,h} \delta (h)_{\rm UHD}  \\
&=  \delta (H^2 h'^2)_{\rm UHD}.
\numberthis
\end{align*}
Plugging this nonadiabatic pressure into the superhorizon conservation equation for $\zeta_h$ \eqref{eq:conservation} we have
\begin{equation}
{\zeta'_h}= -\frac{\delta (H^2 h'^2)_{\rm UHD}}{H^2 h'^2}.
\end{equation}
When the Higgs is slowly rolling, we can use \eqref{eq:hpest} with $c=1$, i.e.\
\begin{equation}
h' \simeq -\frac{V_{,h}}{3 H^2} .
\label{eq:hprime}
\end{equation}
and on uniform Higgs field slicing, the Higgs energy density varies only if $H$ itself varies.
Conversely, ignoring such higher-order corrections in  $\delta \ln H$  (see below), we see that on UHD surfaces $V_{,h}$ is uniform and 
\begin{equation}
{\zeta'_h} = 2 \delta ({ \ln H})_{\rm UHD}.
\end{equation}
Even though the Higgs is a spectator field,
$H$ varies on the UHD slice because of shifts in the number of $e$-folds induced by the
gauge transformation from spatially flat slicing $\delta N =\zeta_h$. We therefore can obtain the desired result
\begin{equation}
\frac{{\zeta'_h}}{\zeta_h} = 2 \frac{\delta \ln H}{\delta N} =-2 \e.
\end{equation}
Using the same logic, we can check our assumption that uniform Higgs field (UHF) and
UHD curvatures coincide to leading order.  Since
\begin{align}
\delta (\rho)_{\rm UHF} = \frac{1}{2}  \delta (H^2 h'^2)_{\rm UHF} = 
-\frac{\epsilon_H}{3}  \rho_h' (\zeta_h)_{\rm UHF} 
\end{align} 
and the gauge transformation between UHF and UHD involves an $e$-fold shift of $\delta (\rho_h)_{\rm UHF}/\rho_h'$, we obtain
\begin{align}
(\zeta_h)_{\rm UHD} \simeq (\zeta_h)_{\rm UHF} \left( 1 + \frac{\epsilon_H}{3} \right).
\end{align}
We therefore use UHF and UHD interchangeably during inflation.  

Finally, notice that these explanations make use of $\delta N$ as computed from gauge transformations.  
We can instead compute it directly in the $\delta N$ formalism using the position-dependent number of $e$-folds $\N_h(h_i;\ h)$ to a UHF slice through the UHF equivalent of Eq.~\eqref{eq:delta_N_linear},
\begin{equation}
\zeta_h = \frac{\pa \N_h(h_i;\  h)}{\pa h_i} \delta h_i.
\end{equation}
To find $\N_h(h_i;\ h)$, we can exploit that the local Higgs evolves along the attractor according to Eq.~(\ref{eq:hprime}), and because the Higgs is a spectator the Hubble rate has no dependence on the Higgs. Working to linear order in the number of $e$-folds $N-N_i$, we expand the denominator $3 H^2$ and we have 
\begin{equation}
h'(N,\vec{x}) \simeq -\left.V_{,h}\right\vert_{h(\vec{x})} \frac{1}{3 H_i^2 (1 - 2 \e (N-N_i) )},
\end{equation}
where $H_i \equiv H(N_i)$. For analytic purposes, we approximate the potential as $V(h)=\frac{1}{4}\lambda h^4$, with $\lambda$ only logarithmically dependent on $h$, solving this equation with $h(N_i,\vec{x}) = h_i $ to find
\begin{equation}
N-N_i = \frac{1 }{2 \e } \bigg(1 - e^{-\dfrac{3 H_i^2 \e}{\lambda} \left(\dfrac{1}{h^2} - \dfrac{1}{h_{i}^2}\right)}\bigg).
\end{equation} 
This is the number of $e$-folds that takes to get from some $h_i$ at time $N_i$ to a field value $h$, and thus we have computed
\begin{equation}
N-N_i = \mathcal{N}_h(h_i;\  h).
\end{equation}
Taking a partial derivative with respect to $h_i$ to get the linear $\zeta_h$, we thus have 
\begin{equation}
\zeta_h = \delta h_i \times \frac{3 H^2}{h_i^3  \lambda} e^{-\dfrac{3 H_i^2 \e}{\lambda} \left(\dfrac{1}{h^2} - \dfrac{1}{h_{i}^2}\right)}.
\end{equation}
To obtain $\zeta_h'$ we take a derivative with respect to the final surface $h$ and find,
\begin{equation}
\frac{{\zeta_h}'}{\zeta_h} = \left. h'\frac{\pa \ln \zeta_h}{\pa h}  = -2 \e \right. ,
\end{equation}
at leading order in $\e$. $h'$ here is evaluated on the background. This gives a third way of understanding the superhorizon evolution~\eqref{eq:sup2}.

We therefore have a consistent picture where the behavior of the flat gauge perturbations, the nonadiabatic pressure, and the $\delta N$ formalism all consistently show that $\zeta_h$ evolves and decays outside the horizon as $H$ evolves during inflation because a uniform Higgs slice is not a uniform Hubble slice.

\subsection{At Reheating}

At reheating, $\zeta_h$ is instantaneously boosted and changes sign. This is due to an instantaneous source in the conservation equation for the Higgs stress-energy due to the direct interaction between radiation and the Higgs, rather than a nonadiabatic pressure. The sign change in $\zeta_h=-\delta \rho_h/\rho'_h$ at reheating occurs because a positive $\delta h$ fluctuation corresponds to a negative $\delta \rho_h$ fluctuation before uplift but a positive one after (see  Eq.~\eqref{eq:drhohafter}). Note that in a realistic inflation model where
$H$ varies smoothly, $\zeta$ would be continuous at reheating.

\subsection{After Reheating}

After reheating, the background Higgs Klein-Gordon equation \eqref{eq:higgs_bg_eom} and our choice of separating out the thermal contributions into a separately conserved $\rho_{\rm r}$ and $\rho_h$ 
leads to the definition
\begin{equation}
\rho_h \equiv \frac{1}{2} \frac{\dot h^2}{a^2} + \VT,
\end{equation}
which implies
\begin{equation}
\label{rhoh_prime}
{\rho}_h' = -3 \frac{\dot{h}^2}{a^2}  -  \VT_{,T} T,
\end{equation}
since the thermal bath redshifts as $T\propto a^{-1}$ to leading order in $\rho_h/\rho_{\rm tot}$.
This last term in the right-hand side of \eqref{rhoh_prime}
was omitted in Refs.~\cite{Espinosa:2017sgp, Gross:2018ivp}.

This separation is equivalent to assuming that the nonthermal displacement of the Higgs field
from inflation oscillates in the effective potential at constant temperature and entropy on timescales short
compared with the expansion.   However,
as emphasized in the main text, our conclusions follow from total energy conservation, which holds independently of this split.

Local energy conservation implies the continuity equation
\begin{equation}
{\rho}_h' + 3  (\rho_h + p_{h, \textrm{eff}}) = 0,
\end{equation}
and by comparison to Eq.~(\ref{rhoh_prime}), we define the effective pressure for the Higgs as
\begin{equation}
p_{h, \textrm{eff}} = \frac{1}{2} \frac{\dot{h}^2}{a^2} - \VT + \VT_{,T} \frac{T}{3}.
\end{equation}
This relation reflects the fact that the effective potential represents thermal components which redshift with the expansion like radiation.

Likewise the Higgs perturbations carry terms associated with the redshifting of the radiation bath.
The flat gauge energy density perturbation is 
\begin{equation}
\label{deltarho}
\delta \rho_h^k = \frac{1}{a^2}\left(\dot{h} \delta \dot{h}^k - \dot{h}^2 A\right) + \VT_{,h} \delta h^k +  \VT_{,T} \delta T^k,
\end{equation}
and the flat gauge effective pressure perturbation is
\begin{align*}
\label{deltap}
\delta p_{h, \textrm{eff}}^k =&\ \frac{1}{a^2}(\dot{h} \delta \dot{h}^k - \dot{h}^2 A) - \VT_{,h} \delta h^k - \frac{2}{3} \ \VT_{,T} \delta T^k \\&+ \frac{1}{3} \left( \VT_{,TT} T \delta T^k +  \VT_{,hT} T \delta h^k\right),\numberthis
\end{align*}
where again the new terms involve the temperature derivatives of $\VT$.
The lapse perturbation $A$ is related to the total momentum density by the 
Einstein constraint equation
\begin{align}
 A ={} & \frac{1}{2} \frac{a H}{k} \frac{\rho_{\rm tot} + p_{\rm tot}}{H^2} (v_{\rm tot}-B),
\end{align}
where the total momentum density,
\begin{equation}
(\rho_{\rm tot} + p_{\rm tot}) (v_{\rm tot}-B) =  \frac{k }{a^2} \dot{h} \delta h^k +
\frac{4}{3} \rho_{\rm r} (v_{\rm r}-B),
\end{equation}
satisfies momentum conservation
\begin{align}
{}& \left[ \frac{d}{d N} + 4 \right]  (\rho_{\rm tot} + p_{\rm tot}) {(v_{\rm tot}-B)}  \nonumber\\
& \qquad\qquad \quad =\frac{k}{aH}\left[ \delta p_{\rm tot}^k + (\rho_{\rm tot} + p_{\rm tot}) A \right],
\end{align}
with $\delta p_{\rm tot}^k = \delta p_{h,\textrm{eff}}^k+ \delta p_{\rm r}^k$ and 
\begin{equation}
\delta p_{\rm r}^k = \frac{\delta \rho_{\rm r}^k}{3} = \frac{4}{3} \zeta_{\rm r}^k \rho_{\rm r}.
\end{equation}
The shift $B$ is then determined from the lapse using the trace-free Einstein equation
\begin{equation}
B' + 2 B = -\frac{k}{aH} A,
\end{equation}
from which we can also see that the expansion shear $-(k/aH)B$ is negligible for
$k/aH \ll 1$ which is required for the $\delta N$ construction of curvature fluctuations to
hold  \cite{Wands:2000dp,Lagos:2019rfc}.  Therefore the initial value for $A$ satisfies the
Hamiltonian constraint
\begin{equation}
A ={}  -\frac{1}{6} \frac{\delta \rho_{\rm tot}^k}{H^2}+ \frac{1}{3} \frac{k}{aH} B 
\simeq  -\frac{1}{6} \frac{\delta \rho_{\rm tot}^k}{H^2},
\end{equation}
for $k/aH \ll 1$.

We then solve the linearized Klein-Gordon equation \eqref{eq:linKGwithmetric} 
with Eq.~\eqref{eq:perturbed_potential} and the total momentum and metric equations above.

Note that this construction assumes any momentum exchange between the Higgs and radiation fields implied 
by the Klein-Gordon equation cancel to conserve the total momentum.  While this exchange itself may not be fully accounted for by the additional effective potential terms, above the horizon all momentum terms are negligible
in their impact on energy density fluctuations and 
\begin{equation}
\delta{\rho}_h' + 3  (\delta\rho_h + \delta p_{h, \textrm{eff}}) \simeq 0
\end{equation}
to leading order in $k/aH$.  
Similarly, though we do not explicitly solve the analogous radiation continuity equation since we assume $\zeta_{\rm r}^k$ is constant,  we have validated that assumption by checking that its  momentum source  is negligible above the horizon.
Moreover, we have also checked numerically that the lapse and shift perturbations have no significant impact on the evolution of $\delta h^k$ or on the final $\zeta$ for superhorizon modes.
This calculation validates the assumptions in the main text.

After solving the Higgs background and perturbation equations, as shown in \S\ref{ssec:linear}, we find that the curvature perturbation $\zeta$ evolves outside the horizon after inflation, as shown in Figs.~\ref{fig:zeta_comparison}--\ref{fig:zeta_nonlin}. This occurs due to a nonadiabatic pressure on a surface of uniform total energy density $\rho_{\rm tot}$,
\begin{equation}
\delta p^{{\rm NA},k}_{\textrm{tot}} = \delta p_{\textrm{tot}}^k- \frac{\dot p_{\textrm{tot}}}{\dot\rho_{\rm tot}} \delta \rho_{\textrm{tot}}^k,
\end{equation}
where all quantities are sums of the Higgs and radiation components.

In \S\ref{ssec:linear}, we saw that $\zeta$ oscillates within each Higgs cycle. The total density perturbation has predominantly canceling components while the adiabatic sound speed $\dot p_{\textrm{tot}}/\dot\rho_\textrm{tot}$ oscillates only fractionally around 1/3. The nonadiabatic pressure is then dominated by the pressure perturbations on the flat slice,
with an oscillatory contribution from the Higgs, explaining the oscillation in $\zeta$.
The oscillations in $\zeta$ increase in amplitude over time, because of the small anharmonic terms in the potential.
Note that curvature oscillations should appear in other related contexts, e.g.\ the curvaton
scenario \cite{Lyth:2002my}, though they are usually averaged over.  

Likewise on the cycle average, at first order the radiation pressure perturbation cancels the Higgs pressure perturbation and $\zeta$ is constant. However, there is a secular drift to $\zeta$ induced by a small noncanceling piece to the cycle-averaged total pressure perturbation. This piece decays as $(\rho_{\rm tot}+p_{\textrm{tot}})$, inducing a constant secular contribution to $\zeta'$ and therefore the small linear drift in $\zeta$ which we discussed in \S\ref{ssec:linear}.
This is the usual entropy fluctuation mechanism to convert isocurvature to curvature
fluctuations through a change in the equation of state of the components 
\cite{Hu:1998gy,Lyth:2002my}, applied here to the
Higgs field.

\bibliographystyle{apsrev4-1}
\bibliography{../references.bib}

\end{document}